%% file: Laycock_astroph.tex
\documentclass[12pt, preprint]{aastex}

\def\degrees{$^{\circ}$}

\newcommand{\lspc}{Lomb-Scargle periodogram}
\newcommand{\lsp}{Lomb-Scargle periodogram }

\newcommand{\Einstein}{{\em Einstein }}
\newcommand{\Einsteinc}{{\em Einstein}}

\newcommand{\rxte}{{\em RXTE }}
\newcommand{\rxtec}{{\em RXTE}} 

\newcommand{\ascac}{{\em ASCA }}
\newcommand{\rosat}{{\em ROSAT }}
\newcommand{\ROSAT}{{\em ROSAT }}
\newcommand{\rosatc}{{\em ROSAT}}

\newcommand{\ASCA}{{\em ASCA }}
\newcommand{\ASCAc}{{\em ASCA}}

\newcommand{\SAXc}{{\em BeppoSAX}}
\newcommand{\Chandra}{{\em Chandra }}

\newcommand{\fluxerg}{\hbox{$\erg\cm^{-2}\s^{-1}\,$}}
\newcommand{\fluxpcu}{\hbox{$\counts\pcu^{-1}\s^{-1}\,$}}

\newcommand{\ergps}{\mbox{$\erg\s^{-1}$}}

\newcommand{\amin}{\hbox{\rm\thinspace arcminute}}
\newcommand{\pcu}{{\rm\thinspace PCU}}
\newcommand{\counts}{{\rm\thinspace counts}} 
\newcommand{\cm}{{\rm\thinspace cm}}

\newcommand{\erg}{{\rm\thinspace erg}}

\newcommand{\s}{{\rm\thinspace s}}

\shorttitle{Pulsars in the SMC}
\shortauthors{Laycock, Corbet, Coe, Marshall, Markwardt, Lochner}

\begin{document}


\title{Long-Term Behavior of X-ray Pulsars in the Small Magellanic Cloud}


\author{S. Laycock\altaffilmark{1,2}, R. H. D. Corbet\altaffilmark{3,4}, M. J. Coe\altaffilmark{1},
F. E. Marshall\altaffilmark{3}, C. Markwardt\altaffilmark{3,5}, J. Lochner\altaffilmark{3,4}}


\altaffiltext{1}{School of Physics \& Astronomy, University of Southampton, Southampton SO17 1BJ, UK}
\altaffiltext{2}{Center for Astrophysics, 60 Garden Street, Cambridge, MA 02138}
\altaffiltext{3}{X-Ray Astrophysics Laboratory, NASA/Goddard Space Flight Center, Greenbelt, MD 20771} 
\altaffiltext{4}{Universities Space Research Association}
\altaffiltext{5}{Department of Astronomy, University of Maryland, College Park, MD 20742}


\begin{abstract}
Results of a 4 year X-ray monitoring campaign of the Small Magellanic Cloud using the Rossi 
X-ray Timing Explorer (\rxtec) are presented. This large dataset makes possible detailed 
investigation of a significant sample of SMC X-ray binaries. 
8 new X-ray pulsars were discovered and a total of 20 different systems 
were detected. Spectral and timing parameters were obtained for 18. 
In the case of 10 pulsars, repeated outbursts were observed, allowing 
determination of candidate orbital periods for these systems. 
We also discuss the spatial, and pulse-period distributions of the SMC pulsars. 
   
\end{abstract}




\section{The Survey}
A program of regular observations of the Small Magellanic Cloud (SMC) using the {\it Rossi X-ray Timing 
Explorer} (RXTE) has been underway since 1997, in conjunction with optical 
imaging and spectroscopic follow-up observations. 
New SMC pulsars discovered by \rxte have been reported in a number of publications, and 
this work is intended to present the wealth of long-term data now accumulated for these sources. 
This paper reports the results of 4 years (November 1997 - February 2002) of weekly monitoring and 
presents the long-term behavior of most of the known X-ray pulsar systems. 
During this time, 16 X-ray pulsars were discovered by \rxtec, at the same time many discoveries were 
made by \ASCA and \SAXc, bringing the number of known X-ray pulsars in the SMC to 30. 

Long term pulsed-flux lightcurves of all pulsars observed during 
the project are presented. 
Twenty pulsars were positively detected in the data presented in this paper, and upper limits 
were placed on 6 other known pulsars that were not detected during this time. In the 
case of 15 systems, multiple X-ray outbursts were observed, raising the possibility of
determining orbital periods. 
Interpretation of the findings, in terms of the overall 
SMC population is discussed. 

\rxte is particularly well suited to monitoring pulsars in the SMC. The  
Proportional Counter Array (PCA) is able to cover a significant fraction of the SMC at sufficient
sensitivity to detect active pulsars with luminosities in the typical range 
10$^{36}$-10$^{38}$ \ergps. 
The spectral response is also favorable as the construction of the
PCA provides its optimum sensitivity across the peak of the typical
pulsar spectrum. The PCA's full-width zero intensity (FWZI) field
of view is 2\degrees\ and the instrument is non-imaging. 
However, the 
PCA can be scanned across a target region and the location of a bright source
determined to an accuracy of 1-2$\arcmin$ by fitting the collimator
response to the observed count-rate. 

The low extinction and unobstructed line of sight to the SMC enable accurate 
measurements of X-ray luminosity and useful observations of the optical 
counterparts with 1-2m telescopes \citep[see for example][]{edge2002}. 
From the X-ray viewpoint, since the distance to the SMC is 
significantly greater than its 'depth' all of the objects
in the SMC can be considered to lie at an equal distance for the purposes 
of luminosity calculations. 


\subsection{Observing Strategy}
The \rxte Proportional Counter Array (PCA) \citep{Jahoda96} was used to make regular observations of 
the SMC to look for new X-ray pulsars and monitor the outbursts of known systems.
The instrument consists of 5 co-aligned Proportional Counter Units (PCU), each has a xenon filled
detector volume with 3 layers of anode-grids for photon detection
plus a lower xenon veto layer for background rejection. Propane-filled veto detectors are positioned
above the PCA to attenuate particle background and non-source direction X-rays.  
Operated in Good Xenon mode, individual photon arrival times are recovered to 1$\mu$s accuracy, 
and pulse-heights processed into 256 channels between 2-100 keV. According to 
the {\it ABC of XTE} the top layer of the PCA is the most sensitive, with the majority of source 
events occurring there, while the background rate is similar in all 3 layers. We verified that 
the detection significance for pulsars is slightly better using layer 1 only versus layers 1 and 
2, when extracting lightcurves in the 2-10 keV band.

\begin{figure}
\includegraphics[width=10cm,angle=-90]{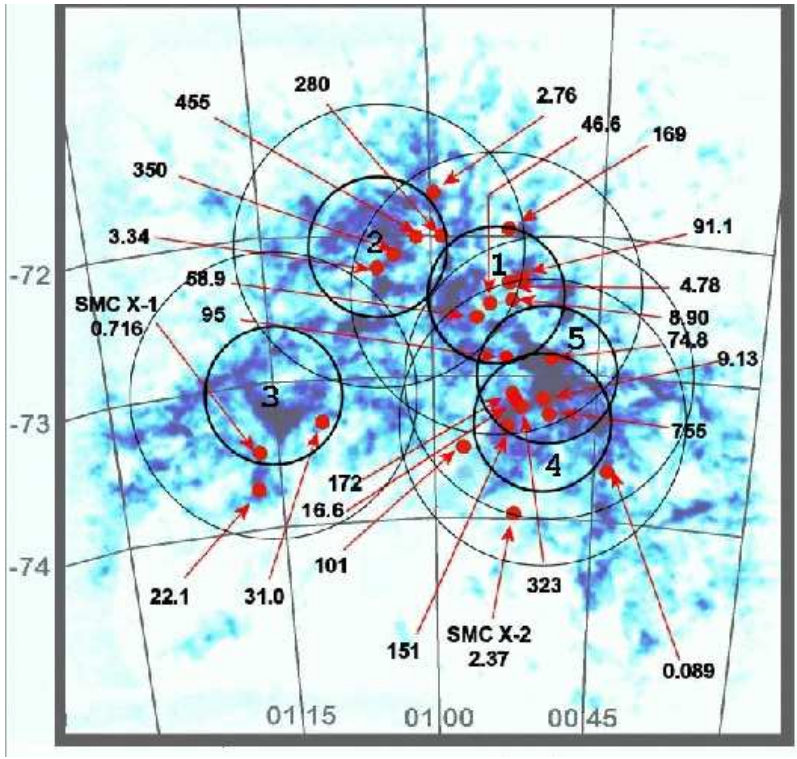}
\caption{\rxte observations in the SMC. Circles show the PCA field of view in the most-used 
monitoring positions. The pointing centers are listed in Table~\ref{tab:positions}. The background 
image is an HI radio map by \citet{S99}. The positions of known pulsars are indicated by their 
pulse periods.}
\label{fig:smcpic}
\end{figure}

The program began after a new X-ray transient was detected during a PCA slew
in the vicinity of SMC X-3 \citep*{IAU6777}. A PCA observation was
conducted on 1997 November 29 in response to this detection, revealing 
not SMC X-3 but three previously unknown pulsars with pulse periods of 91.s, 74s and 
46.6s \citep{IAU6803}. Continued observations throughout December and
January detected 2 more pulsars with periods of 169s \citep{IAU6814}
and 58.9s \citep{IAU6818}. At times observations were being made at a
rate of 1 per day, giving excellent temporal coverage of the outbursts of 
these pulsars. Following these observations, a program of 
regular monitoring was established (see Table~\ref{tab:positions}), with
pointed PCA observations of several ksec duration made on a weekly basis. 
Pointing positions were carefully chosen to give good coverage of the main body of the SMC, 
as shown in Figure~\ref{fig:smcpic}. 
The ``wing'' was not regularly observed due to the presence of the luminous
supergiant High-Mass X-ray Binary (HMXB) SMC X-1.
This source was deliberately excluded from the PCA survey because its persistent emission, very 
strong pulsations, and timing noise mask the presence of other weaker pulsars when it is in the 
field of view. Observations were made in 3 phases during which slightly different 
observing strategies have been employed, as described below. 

\emph{Phase 1.}
This comprises the series of observations conducted and described by \citet{Lochner99a}. The observations 
were made primarily in two positions designated 1a and 1c.
Position 1a was centered on the location of SMC X-3, because the initial source detected 
was at first thought to be SMC X-3. 
Position 1b was the position of a single PCA pointing 
aimed at the location of one of the new pulsars.
Position 1c lies close by and was used throughout 1998 to 
monitor the activity of the 4 newly discovered pulsars mentioned above.  
 
\emph{Phase 2.} 
During this period we continued to monitor the known pulsars and at the 
same time search for new sources. Position 1 largely overlaps with the 
fields of view of the earlier observations. Position 3 was chosen to cover 
the eastern wing of the SMC which contains SMC X-1. Positions 2 and 4 were 
chosen to cover the rest of the SMC.

\emph{Phase 3.}
After 2 years of monitoring position 1, we broadened the search by shifting south 
to position 5, also in the main body of the SMC. 
The supplementary positions 2, 3 and 4 were not observed, as previous observations of these 
positions caused gaps in coverage of the main position. 
Position 5 was observed each week for 2.1 years.

\subsection{Data Reduction Pipeline}
Data reduction was performed in two stages, ``real-time monitoring'' and ``survey'':
Immediately after each routine PCA observation, quick-look data were searched for new pulsars. 
Initially a lightcurve was generated at 3-10 keV and 0.1 second time resolution. No 
filtering criteria or background subtraction were applied at this stage. The power spectrum 
was inspected visually and with a peak search algorithm. This preliminary data analysis
was usually performed within a day or so of the observation taking place, to identify 
new pulsars and schedule rapid TOO follow-up observations. 
The long-term survey made use of the production data, and more detailed analysis methods.
All data reduction was done with {\it Ftools}. Standard filtering criteria 
were applied to exclude data 
collected during: periods of high background;
times when the source was attenuated by Earth's atmosphere; during slews on/off source; and times 
immediately after SAA passage. On 2000 May 20 the upper propane veto-layer of PCU 0
was lost, after that date PCU 0 data were excluded. Filtering by detector/layer was performed 
with bit-masks generated by {\it sefilter}.

Background count-rates were generated using {\it pcabackest} and the L7 models for faint 
sources. Lightcurves and spectra were extracted from the generated background using {\it saextrct} 
with the same filtering criteria and detector/layer combinations as were applied to the science data.
 
\subsubsection{Lightcurve Extraction}
Science lightcurves were extracted using {\it seextrct} at 0.01 second time binning for the 3-10 keV 
Good Xenon data from the top anode layer of each active PCU. This configuration maximizes 
signal-to-noise for pulsars. Background lightcurves were extracted with 16s binning (the 
minimum available) and subtracted from the science lightcurve, without modifying the errors 
on the science count-rates. 

The lightcurve was then normalized to \fluxpcu, by dividing each flux value by the number of PCUs 
active at that time. Lightcurve bin-times were corrected for the motion of the satellite and 
Earth's orbit using {\it faxbary} to give the arrival times at the solar system barycenter. 

\subsubsection{Correcting for Collimator Response}
For every pulsar with a well known position, the detector sensitivity ($R$) was calculated at each 
pointing position. 
The collimator for each PCU are constructed from corrugated
sheets with hexagonal cells. The hexagonal cells have small
offsets in pointing direction and the angular responses for each PCU
are different. For the purposes of examining long term light
curves we adopt a simplified model for the collimators of
a circular field of view with a simple triangular collimator 
response of full-width half maximum (FWHM) = 1\degr, full-width
zero intensity (FWZI) = 2\degr.
This simplified model differs from the true collimator response
in being somewhat sharper peaked, lacking extended wings at larger
angles (where we do not have significant pulsar detections) and
lacking the hexagonal symmetry which is also most pronounced
at larger offset angles. The use of this simplified model
speeds data analysis and is not a significant factor
in, for example, searching for periodic behavior.
With our simplified collimator model, for a particular pulsar at known distance 
from the pointing center, $R$ equals 1.0 minus the net pointing offset in degrees.
Values of {\em R} for each pulsar in each of the monitoring positions 
are listed in Table~\ref{tab:responses}. This table should be consulted when examining the 
long-term pulsar lightcurves in Section~\ref{sect:lcurves}. It is apparent that the scatter in flux 
measurements is correlated with $R$. 
Blank entries in Table~\ref{tab:responses} indicate that the pulsar was not in the field of view. 
In practice, observations at marginal sensitivity were also excluded. A value of $R < 0.1$ was 
regarded as marginal and a lightcurve not extracted for observations of that particular
pulsar/pointing position combination.  For spectral analysis in this paper we utilize the full
detailed model of collimator response for those sources with accurate
positions.
 
\subsection{Timing Analysis}
The methods used to measure pulse periods, pulse amplitudes, calculate significance levels and 
reject false detections are outlined here. 

A \lsp \citep{Scargle82} was calculated for each observation, scanning the range 0.02-1000 s at a 
resolution of $1\times 10^{-5}$ Hz. Also generated by the pipeline were statistical parameters 
used in assigning significance levels and converting between spectral power and pulsed flux.

The special properties of the \lsp were exploited to scan every observation around the period 
of each known pulsar. 
We first determined if there was $significant$ power at a given (known pulsar) frequency and then 
determined its period and amplitude, if however there was non-zero power at a low significance 
level, an upper limit was placed on the pulse-amplitude of the (presumably inactive) pulsar. 

For each known pulsar, (i.e those detected at a high level of confidence in one or more 
observations, or from the literature) an appropriate period search-range was determined 
from inspection of occurrences detected at $\geq$ 90\% significance in a \emph{blind search}. 
For ``literature'' pulsars a $\pm$5\% frequency band centered on the known period was used.
The \lsp of each observation was then scanned in this period 
range, the maximum power identified and its significance estimated, considering only the 
independent frequencies in the allowed period range. This is the \emph{prior knowledge} 
significance.

Given that 30 SMC pulsars are known with periods less than 1000 s, 
allowing for up to 3 harmonics gives a possible 93 genuine signal frequencies. Once period 
variability and timing resolution are included, the possibility of contamination requires the 
adoption of an algorithm to reject false detections of pulsars that lie close to the harmonic 
frequencies of others. 

If the maximum power recorded was not a \emph{peak} inside the selected range, 
i.e. it lay at edge of the frequency range, then it was 
assumed to be leakage of power from a nearby and unrelated frequency. Whenever this situation 
occurred, the detection significance was set to zero since the probability that the power was 
due to the pulsar being searched for is negligible. For positive detections the pulse period and 
its uncertainty were also recorded. The uncertainty on the period was calculated using 
the formulation of \cite{kovacs1980} which accounts for the strength of the detected signal.

For each pulsar we only considered observations of duration greater than 4
pulsation cycles otherwise there are unacceptably large errors on the
pulse period and poor confidence in their identification.

For each positive detection, the measured power $P_{LS}$ was converted into a pulse amplitude using 
Equation~\ref{eqn:pow2amp}: where the lightcurve has $N_P$ points and variance $\sigma^2$, the 
peak-to-valley amplitude (twice the sinusoidal amplitude) $A^{\ast}$ of a signal detected in the \lsp is: 
\newline 
\begin{equation}
\label{eqn:pow2amp}
A^{\ast} = 4\sqrt {\frac{P_{LS} \sigma ^2}{N_P }} 
\end{equation}

The significance of peaks in the \lsp is described by a simple exponential function, related to 
the number of independent frequencies in the range being searched in the case of periodic 
signals superimposed on white noise with a Gaussian distribution. The significance calculations in 
this work follow the prescription of \citet{press93}. 

The precise nature of the data sampling is relevant in timing analysis. Data gaps are unavoidable 
and applying barycenter correction to the lightcurve bin times causes a small departure from 
regular sampling. Random sampling is $recommended$ by \citet{Scargle82} in order to 
maximize the anti-aliasing properties of the Lomb-Scargle periodogram, but clumpy or gappy 
data are not handled so efficiently and, in cases involving one or two 
large gaps in otherwise regular data, aliasing is similar to a Discrete Fourier Transform. We used 
the fast coding of \citet{press89}. 

\subsection{Flux Measurements}
\label{section:flux}
Throughout this paper we use units of \fluxpcu, these can be 
approximately compared to the flux in \fluxerg if assumptions are made regarding the X-ray 
photon spectral index ($\alpha$) and line-of-sight absorption column-density ($N_H$). 
Taking reasonable values for X-ray pulsars to be $\alpha=1, N_H=1\times10^{22}cm^{-2}$ gives an 
approximate conversion factor of 1\fluxpcu = 1.2$\times10^{-11}$\fluxerg. At the SMC (65 kpc), 
this corresponds to a luminosity of $L_{X}=5.7\times 10^{36}\ erg\ s^{-1}$.
Pulse amplitude is simply related to the pulsed flux (pulsed component summed over a full cycle)
by a factor of 2. The total flux can only be determined if the pulse fraction (ratio of pulsed 
to unpulsed component) is known. We choose to discuss the pulse amplitude as this is the 
most directly measurable quantity and enables robust measurements to be
made even when multiple pulsars are present in a single observation.

\subsection{Orbital Period Estimation}
\label{sect:sta}
One of our primary goals was to measure orbital periods for the X-ray sources. This was done 
by analyzing the long-term pulsed-flux lightcurves produced by our pipeline. In prototypical 
Be-X systems, X-ray outbursts occur in periodic sequences. Many of our lightcurves
presented in Section~\ref{sect:lcurves} show such features, for example XTE J0055-724 
(Figure~\ref{fig:59amp}). In these cases we employed Phase Dispersion Minimization 
\citep[PDM,][]{Stellingwerf78}, which is the most appropriate technique to search for highly 
irregular modulations which are also poorly sampled. 

Systems such as AX J0049.4-7323 (Figure~\ref{fig:755amp}) presented two outbursts which, 
although insufficient to formally claim a ``period'', have nonetheless proved to be highly reliable 
indicators of the orbital period. If two outbursts are seen from the same Be system within a 
recurrence time of several weeks, we know from the underlying physics that these are highly
likely to be separated by one orbital period. Measurement of such ``recurrence times''
was done by folding the dates of significant detections on a range of trial periods, and selecting the 
period that produced the minimum scatter in phase. This is similar to taking the mean separation
of outbursts, but includes multiple data-points for each outburst, and accounts for the fact that 
we do not know the date of outburst-maximum very 
well, due to sparse sampling. The uncertainty is calculated by multiplying the minimum standard 
deviation in phase by the best period. We term this method ``simple timing analysis'' (STA). Given 
the small number of points available (typically 4-20), STA results are a ``best estimate" only.

In other cases (e.g. AX J 0051.6-7311, see Figure~\ref{fig:172amp}) the long-term lightcurve
contains many detections of the pulsar on dates that do not form an obvious periodic sequence.
The explanations include X-ray emission that is genuinely aperiodic, as well as missed outbursts which
could hide an underlying periodic pattern. These cases were analyzed with both PDM and STA and 
the most likely periods or recurrence times reported, in order to provide a guide for other observers.
 
Finally a few systems (e.g. SMC X-2) have undergone a single giant outburst during our
monitoring project. In these cases no orbital periods could be  measured, instead pulse-period 
variations were used to constrain the contributions from orbital motion (Doppler shifts) and 
accretion torques.

\section{Results I: Pulsar Monitor Charts}
\label{sect:monitor}
After reduction, the data were analyzed in 2 separate ways to search for variable sources, 
the first of these is the blind search.

\begin{figure*}
\begin{center}
\includegraphics[width=15cm]{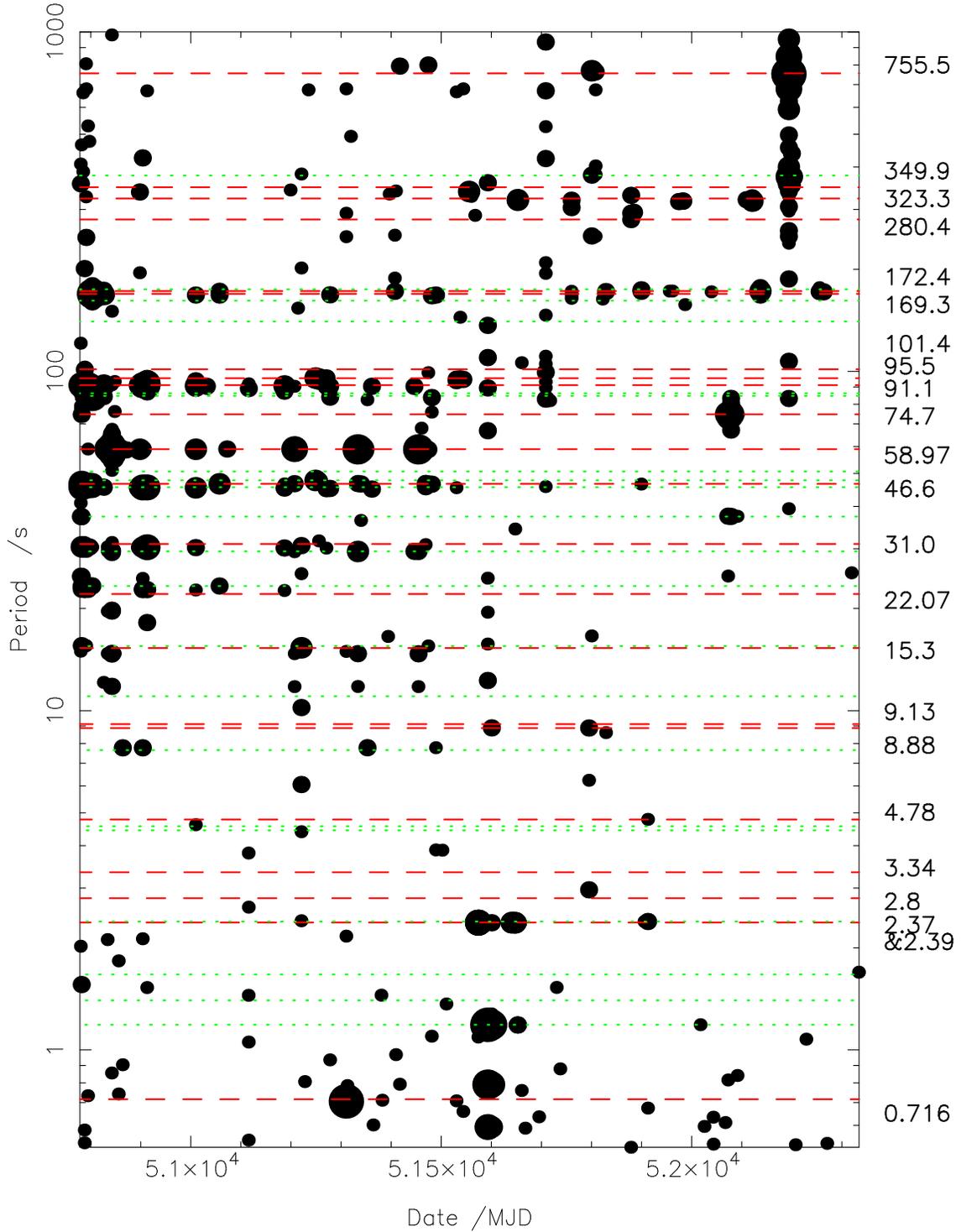}
\caption{Activity chart for pulsars detected in all of the regular pointing positions between
1997 November 27 and 2002 February 28. Relative pulsation strength is indicated by marker size.
No selection criteria have been applied other than a significance threshold. All power at greater   
than 90\% blind search significance is plotted.}
\label{fig:peaks12345}
\end{center}
\end{figure*}

Having generated the \lsp for every observation, the resulting database of periodograms was 
searched for significant peaks. The significance level for each observation was estimated 
individually. This stage was executed as a ``blind search'', 
the threshold adopted was 90\% significance considering all frequencies in the (period) range 0.5 
-1000 seconds.  We used a simple boxcar peak-search algorithm which identified a peak as
any frequency having a greater power than its nearest neighbors, subject to that power being
above the 90\% significance level. 
Figure~\ref{fig:peaks12345} displays the combined results of this analysis in the form 
of a ``pulsar activity monitor'', showing the activity of bright pulsars in all the 
monitoring positions. 
The size of the graph markers indicates spectral power on a logarithmic scale, horizontal 
dashed lines (red) indicate pulse periods of the majority of known 
pulsars in the SMC, these periods are given on the right of the plot. 
In addition to detecting pulse periods, this procedure also picks up harmonics and, in 
principle at least, low frequency quasi-periodic oscillations. 
In particular it has been observed in the course of this work that at times the $P_{pulse}/2$ 
harmonic can dominate the power spectrum. Therefore all significant power is included on the plots.
The expected $P_{pulse}/2$ harmonics of known pulsars are indicated by dotted (green) lines.
As the various pointing positions cover partially overlapping fields of view, different pulsars 
are visible in subsets of the data. Positions 1 and 5 comprise the bulk of the data. 

Figure~\ref{fig:peaks12345} shows pulsars ranging in period 
from 0.7s (SMC X-1) to 755s (AX J0049.4-7323). It is immediately apparent that two sources 
were particularly active throughout the survey. 
The 172.4s and 323s pulsars appear to have undergone respectively 5 and 7 
outbursts over a period of about 700 days. If these outbursts are normal Be/X-ray binary outbursts 
then the results are suggestive of orbital periods of (very) approximately 700/5=140d and 700/7=100d.
The line labeled ``2.37 \& 2.39'' is also of particular interest as the points lying along it 
actually belong to two different pulsars. The group of points at about MJD 51600 is due to an 
outburst from SMC X-2, Below these points, 2$^{nd}$, 3$^{rd}$ and 4$^{th}$ harmonics are visible. 
There then follows a short ($\sim$ 2 week) break in observing coverage after which the 
pulsations were again detected but visibly weaker and with the relative strength of the harmonics 
altered. Just after MJD 51900 a third group of points appear on the ``2.37 \& 2.39'' line, 
these are the first harmonic of XTE J0052-723 which was first discovered in these observations, 
the fundamental is seen just once on this plot. 

No sub-second pulsars were detected, only 3 observations containing significant peaks at P$<$1s were 
ever seen:

(1) A single detection of the (P/2) harmonic
of SMC X-1; and 2 harmonics of SMC X-2 (P/5 and P/6). 

(2) On MJD 51381, period 0.0329429(5)s, amplitude 0.73 \fluxpcu, significance above 99\% in one 1050s 
observation. However two other observations on the same day, of comparable length made $~$1 hr before and $~$1 hr 
after failed to show any evidence for this period. 
 
(3) On MJD 51432, period 0.0747731(3)s, amplitude 0.31 \fluxpcu, significance 97\%.
It is expected that a small number of false detections will occur given the large number of observations 
analyzed and this low significance detection may thus not be real.

\section{Results II: Long Term Pulsar Lightcurves}
\label{sect:lcurves}
Long term behavior of all known pulsars in the SMC was investigated
using the database of \lspc s generated from the \rxte monitoring observations. 

In the following section the pulsars are ordered by pulse period. Trends in physical characteristics 
are likely to follow the pulse period, so it is appropriate to list the systems in a natural 
sequence that facilitates comparison. 

The most important results are shown in a series of 3-panel plots for each source, laid out as 
follows:

\noindent {\it Top panel.} Pulsed flux lightcurve, filled symbols indicate positive detections, 
open symbols are upper limits.
\newline {\it Middle panel.} Pulse period with uncertainty.
\newline {\it Lower Panel.} Statistical significance of each pulsed flux measurement. 
Two significance estimators are plotted: Blind search (squares), and prior 
knowledge of period (circles). The criterion for a positive detection was 99\% prior knowledge 
significance. Blind search significance values are shown in order to convey those detections of 
outstanding magnitude.


\subsection{SMC X-2 (2.37s)}
SMC X-2 was discovered by \emph{SAS-3} in 1978 \citep{Clarke78} and although outbursts were 
also observed by \emph{HEAO 1} and \rosatc, pulsations had not been detected until the \rxte 
monitoring data presented here.

A single outburst was detected from SMC X-2, lasting from 2000 January 24 to 
April 23. During this outburst, the luminosity (2-25 keV) reached a peak of 
4.7 $\times 10^{38}\ergps$, and was detected down to a level of 5.7$\times 10^{37}\ergps$ 
before disappearing from view sometime before May 5.
The detection of SMC X-2 with the \rxte All-Sky Monitor (\emph{ASM}) followed by discovery of 2.37 second 
pulsations co-incident with the known position of SMC X-2 from \emph{PCA} scans and targeted 
observations was presented by \citet{Corbet01_x2}, along with contemporaneous optical observations 
confirming the counterpart originally proposed by \citet{Murdin79}. An \ASCA observation on April 
24 by \citet{Yokogawa01} confirmed the 2.37 s pulsar to be exactly coincident with the position 
of SMC X-2 determined by \emph{SAS-3} and \rosat.  

\begin{figure}
\includegraphics[width=10cm,angle=-90]{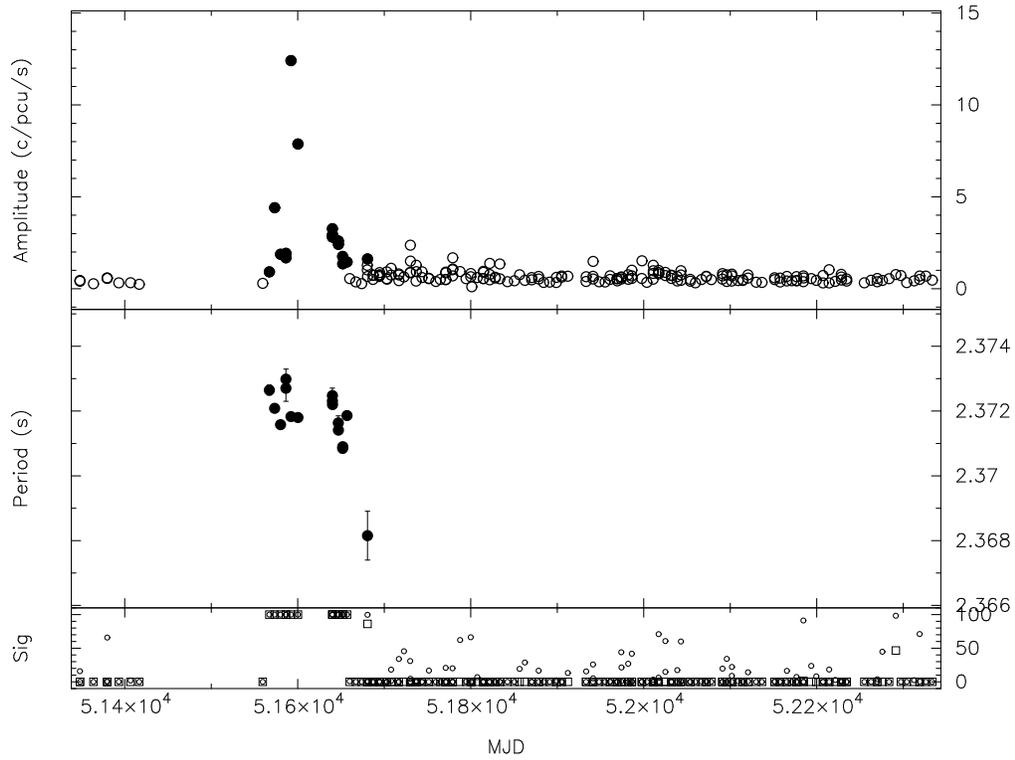}
\caption{SMC X-2 pulsed flux history. Top panel: pulse amplitude in \fluxpcu, filled 
symbols indicate $>$99\% detection significance. Middle panel: pulse period and uncertainty for 
positive detections. Bottom panel: detection 
significance, squares= blind search, circles= assuming prior knowledge of the pulse period.}  
\label{fig:smcx2amp}
\end{figure}

From the SMC monitoring data, pulsations with a period of 2.37s were detected with the PCA during 
13 observations between MJD 51567 and MJD 51657. Of these, 9 were pointings at the regular 
monitoring position (position 5, see Table~\ref{tab:positions}), plus 2 targeted pointings centered 
on SMC X-2, and 2 sets of scans across the source region. All of these observations are included 
in the long-term lightcurve (Figure~\ref{fig:smcx2amp}).

The duration of the outburst was constrained by non-detections before and after the dates given 
in Table~\ref{tab:x2flux}, on MJD 51560 and 51666-7 and 51670. The upper limits for the end of 
the outburst are more stringent as these observations were pointed directly at the position of 
SMC X-2.

\begin{figure}
\includegraphics[width=10cm,angle=-90]{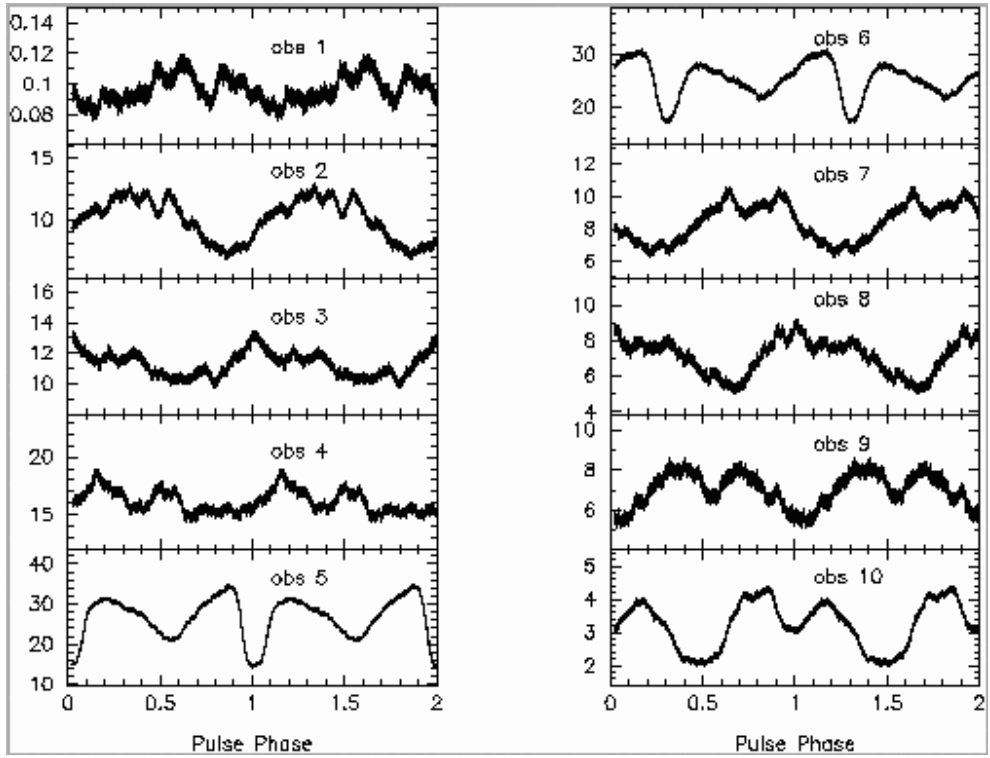}
\caption{Pulse profiles of SMC X-2 during the January - April 2000
 outburst. Flux in units of \fluxpcu for the 2 - 10 keV band, arbitrary phase.
 Collimator correction and background subtraction have been applied.}
\label{fig:x2pro}
\end{figure}

Due to changes is the power spectrum of SMC X-2 during the outburst, pulse periods 
were refined by PDM \citep{Stellingwerf78}. 
The refined periods are given in Table~\ref{tab:x2flux}. 
The pulse profiles (Figure~\ref{fig:x2pro}) obtained during the outburst show an 
interesting trend with luminosity. During the low luminosity observations at the beginning and end 
of the outburst, the pulse profile was a weakly double peaked shape, with roughly equal peaks 
separated by half a cycle: typical of that seen in many X-ray pulsars. At 
high luminosity the shape changed, with peak flux occurring either side of a narrow minimum, 
the separation between the first and second peak now 0.7 in phase. This change was seen in the 
power spectra as a reversal in the normal ratio of the fundamental and 1st harmonic: 
evident from the pulsar monitor (Figure~\ref{fig:peaks12345}) 
around MJD 51600. 
For this reason the pulse amplitudes plotted in Figure~\ref{fig:smcx2amp} are derived either 
from the fundamental or harmonic depending on which was the stronger.

The X-ray spectrum of SMC X-2 was extracted for 3 observations, the two targeted pointings
\citep{Corbet01_x2} and observation 5 when the highest flux was observed. 
Assuming the spectral parameters in Table~\ref{tab:x2spec}
to be reasonably representative of SMC X-2 during the whole 
outburst ($N_H=2\times10^{22}$ cm$^{-2}$, $\alpha=1$) the fluxes reported 
in Table~\ref{tab:x2flux} may be converted to $L_X$ by $1\fluxpcu = 1.72\times10^{37}\ergps$   
assuming a 65 kpc distance to the SMC.

\subsubsection{In-outburst Timing Behavior}
Period variations evident in Figure~\ref{fig:smcx2amp} are suggestive of spin-up or orbital modulation, 
these measurements were investigated and refined by Phase 
Dispersion Minimization (PDM). The PDM method
was used because the shape of the pulse profiles are highly irregular 
and change between observations. A period range was selected that encompasses all the periods 
determined from the power spectra, hence a range of 2.371-2.373s at a resolution of 10$^{-6}$\ s 
rather than the fundamental. These periods are given in Table~\ref{tab:x2flux}.
The magnitude of the variations is 1.7$\times10^{-3}$\ s, significantly larger than the mean uncertainty 
(3.9$\times10^{-5}$\ s) in the period determination. The actual timing behavior is somewhat 
surprising as there appear to be two spin-up events on similar timescales, one before the gap in 
coverage and one after. Observations 4 and 10 seem not to fit this picture, which may be attributable 
to systematic errors resulting from pulse profile variation. 

A problem with the spin-up interpretation is the size
of the pulse period changes. Standard accretion torque theory \citep{GL1979} predicts spin period 
changes of approximately 2.4 $\times$10$^{-11}$ s s$^{-1}$ L$_{38}^{6/7}\mu_{30}^{2/7}$ which is exceeded by factors 
of up to $\sim$100 (Table 2).
Another contribution to period changes can come from orbital motion.
For a circular orbit, and assuming a neutron star mass of 1.4\ M$_{\sun}$ and
a mass donor mass of 10\ M$_{\sun}$ then the orbital velocity semi-amplitude
is $\sim$170 (15/P$_{orb}$)$^{1/3}$ sin i km s$^{-1}$.
where P$_{orb}$ is in units of days.
The observed period changes are of at least of this approximate magnitude
although an obvious systematic trend was not detected. Finally, if our
period determinations are dependent on pulse profile, although we do
not expect this, then an artificial variation of pulse period with
luminosity may be produced.
   
The pulse period discovered for SMC X-2 is among the shortest known for HMXBs especially for 
those containing a Be star primary. If SMC X-2 follows the loose relation 
between $P_{pulse}$ and $P_{orbit}$ seen in Be/X-ray binaries \citep{Corbet86} then its orbital period may be around 15 days. 
The pulse profiles appear to be correlated with $L_{X}$, similar changes were observed in 
XTE J0052-723 and also the 51 s XTE pulsar that was initially reported as having a period 
of 25.5s by \citet{lamb2001}. The temporal coverage of the \rxte observations of 
SMC X-2 presented here were sufficient to track the luminosity and pulse period behavior 
reasonably closely for the duration of an entire outburst. A gap in coverage occurred between 
MJD 51600-51640 and certain observed properties suggest that 2 outbursts may have occurred. 
Evidence for this interpretation comes from the flux history and the pulse period variations. 
If two separate outbursts did in fact occur, then an estimate can be made that the orbital 
period is less than approximately 70 days, by taking the difference of the dates on which the 
two periods of spin-up began. 


\subsection{XTE J0052-723 (4.78s)}

\begin{figure}
\includegraphics[width=10cm,angle=-90]{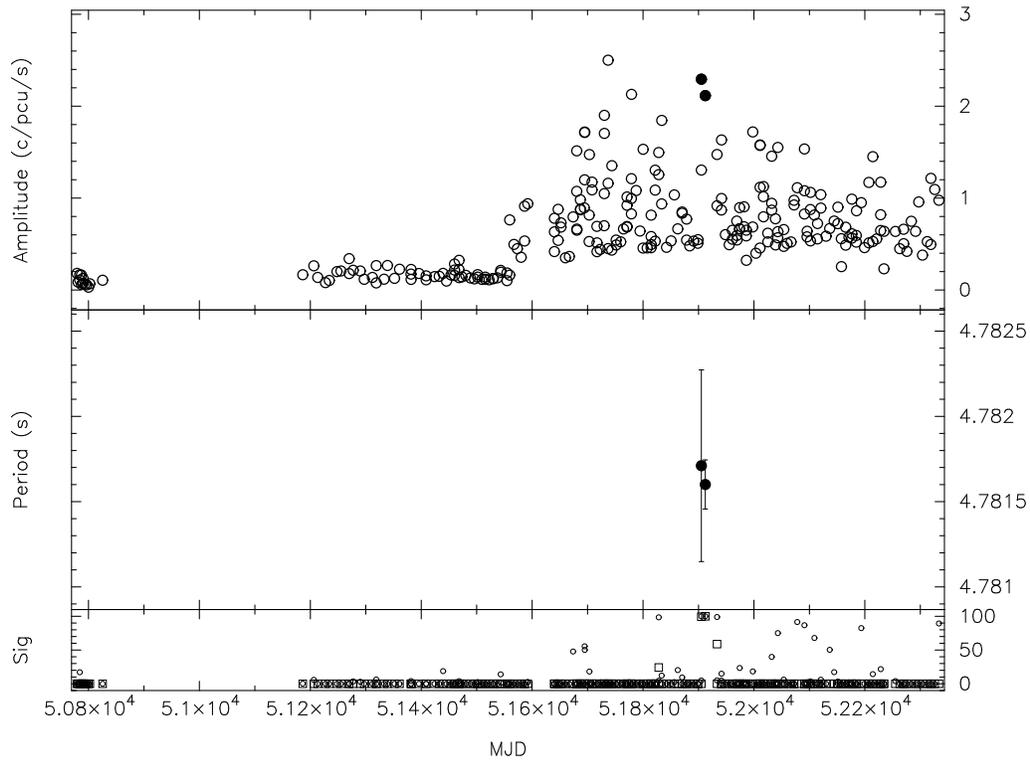}
\caption{XTE J0052-723 pulsed flux history. Top panel: pulse amplitude in \fluxpcu, 
filled symbols indicate $>$99\% detection significance. Middle panel: pulse period and 
uncertainty for positive detections. Bottom panel: detection significance, squares= blind 
search, circles= assuming prior knowledge of the pulse period.}   
\label{fig:4.78amp}
\end{figure}

Pulsations with a 4.78 s period were detected on 
2000 December 27 during an observation of pointing position 5. 
An analysis of the X-ray and optical observations was presented by \citet{Laycock02b}
identifying a possible B0V-B1V counterpart. 
No additional outbursts were detected, as evidenced by Figure~\ref{fig:4.78amp}. 
For most of the outburst the pulse profile was strongly double peaked, causing the $P_{pulse}/2$ 
harmonic power to be more indicative of the actual pulsed flux, a feature that is evident in the pulsar monitor 
(Figure~\ref{fig:peaks12345}). In all cases around the time of the outburst, the amplitude 
of the $P/2$ harmonic in Figure~\ref{fig:4.78amp} was plotted if it was greater than the fundamental. 

\subsection{2E 0050.1-7247 (8.88s)}
The pulsar activity monitor (Figure~\ref{fig:peaks12345}) shows two strong detections 
of a $\sim$9 second pulsar which appear to correspond to
the 8.88s pulsar 2E 0050.1-7247. The pulsed flux history reveals a number of detections
in 3 groups separated by $\approx$ 200 days.
Although some of the detections appear to at least
roughly coincide with detections of the 16.6s pulsar the uncertainties
on our period determinations apparently exclude the possibility
that pulsations at 8.88s are harmonics of the 16.6s source. A 2-10 keV pulse profile
obtained during the brightest detection of 8.88s pulsations is shown in
Figure~\ref{fig:8.88pro}, it is triple peaked.


\subsection{RX J0052.1-7319 (15s)}
This pulsar was not conclusively detected in any regular monitoring observation,
but pulsations with a 15.7s period were detected in a special deep observation described in 
Section~\ref{sect:deep}.
The source was very faint and only detected due to the length of the observation,
pulse amplitude was 0.12 \fluxpcu. 
No spectrum could be obtained due the many active pulsars in the field of view, including 
the close period of 16.6s at about 5 times greater amplitude.


\subsection{XTE 16.6 seconds}
The 16.6 second pulsar was discovered in a deep observation of position 4 
(see Section~\ref{sect:deep}).
The pulsar appears in Figure~\ref{fig:16amp} on 8 occasions, which seem to belong to 6 
separate outbursts. Simple timing analysis (Section~\ref{sect:sta}) was performed on the
99\% detections, suggesting a candidate orbital period of 189$\pm18$ days, with $T_0$ = MJD 51393. 
The folded lightcurve is 
uninteresting as the fluxes when the pulsar is detected are similar to the upper limits
in non-detections. The 2-10 keV pulse profile Figure~\ref{fig:16pro} is approximately sinusoidal.

An analysis of archival \rxtec, \ROSAT and \ASCA data has now appeared in the 
literature \citep{lamb2001} and a tentative association made with the \ROSAT source  
RX J0051.8-7310 on the basis of a marginal detection of periodicity in data from \ROSAT and \ASCAc. 
\citet{Yokogawa02} demonstrate this identification is incorrect. From our observations the 
position of the 16.6 second pulsar is constrained to lie within the overlap of the PCA field of 
view at positions 4 and 5. 


\subsection{XTE J0111.2-7317 (31s)}
This source was in the survey field of view on 3 occasions. 31 s pulsations were detected in 
only the first of these observations on MJD 51220. This detection coincides with the very end 
of a giant outburst of this system which was simultaneously discovered 
by \rxte and BATSE \citep{Chakrabarty98a,WF98}. 
For the single PCA detection, the pulse period was 30.65$\pm$0.05 s The spectral fit to this 
observation gave an unabsorbed 2-10 keV luminosity of 4.6$\times$10$^{37}$ \ergps, and showed a 
prominent 6.4 keV iron line, full spectral parameters given in Table~\ref{tab:catalogue}. The 2-10 keV 
pulse profile for this observation is shown in Figure~\ref{fig:31pro}, it is highly irregular, 
featuring three distinct peaks and one deep minimum per cycle.

\subsection{1WGA J0053.8-7226 (46.6s)}
The 46.6 second pulsar was one of three sources discovered in the vicinity of SMC X-3 in 
1997 November \citep{IAU6803} and remained active through December. Its subsequent 
reappearance on 1998 August 3 was reported by \citet{IAU7007} who suggested an orbital 
period of 139 days based on these 2 outbursts.
After analyzing the long-term monitoring data, pulsations at 46.6 s were positively 
detected in 54 observations, apparently grouped in 8 separate outbursts. The full dataset 
presented in Figure~\ref{fig:46.6amp} was used to determine the orbital period of the system. 
This was done in two stages because the data quality was not constant due to changes in 
pointing position. For observations made at positions 1a, 1b, 1c and 1 the pulsar was at or 
close to the center of the PCA field of view. This subset of the data was analyzed using PDM, 
only one likely period was discovered 
at 137$\pm$8 days. In order to include the two later outbursts, the position 5 data were 
filtered to remove all points at less than 99\% significance and those remaining were added to 
the first dataset and reanalyzed. This procedure was justified because the source was poorly 
placed in the position 5 field of view and probably only detectable close to the peak 
of the last 2 outbursts. The addition of these points slightly deepened and narrowed the PDM 
minimum  to give a period of 139$\pm$6 days, the uncertainty is the FWHM of the PDM minimum. 
Numbering orbital cycles from the first outburst, the observed detections correspond to phase 
zero of cycles 1, 2, 3, 4, 5, 6, 9 and 12, where the adopted zero-point is MJD 50779. The folded 
lightcurve is shown in Figure~\ref{fig:46fold}.

Two candidates have been proposed for the optical counterpart \citep{Buckley01}, both 
lie in the error box determined by \ROSAT and \ASCA and both show strong H$\alpha$ emission 
lines and photometric colors of Be stars. One of the candidates was also reported to exhibit 
an IR excess and variability. 


\subsection{XTE J0055-724 (59s)}
Strong pulsations at a period of 59 s were discovered by \citet{IAU6818} during a 
search for pulsars in the vicinity of SMC X-3. 
The 59 s pulsar appears to have been emitting approximately regular 
outbursts throughout the lifetime of \rxte. The pulsar was in our field of view for 
the entire monitoring program although it was only detected in positions 1, 1a, 1b and 1c. 
XTE J0055-724 was close to the center of the field of view in these observations but only 
marginally covered by the other positions, as a consequence the upper limits for the pulsed 
flux are rather large after MJD 51555. Figure~\ref{fig:59amp} shows just the 
position 1, 1a, 1b, and 1c results, only this subset of the data was used for determining the
orbital period. The 2-10 keV pulse profile (Figure~\ref{fig:59pro}) is dominated by a single 
asymmetric peak with some finer structure visible.

Four separate outbursts were observed, each reaching a similar brightness and occurring on 
similar timescales.
Averaging over the latter 3 peak fluxes gives a mean of 2.0 \fluxpcu with a spread of 0.1 
\fluxpcu. The timescale for the flux to go from an approximately quiescent level, up to peak, 
and back down again was about 40$\pm$5 days. The first outburst had excellent temporal coverage 
and was possibly brighter than the other 3, although this may just be because more 
observations were made during the peak of the outburst.

A timing analysis was conducted to identify a possible orbital period in XTE J0055-724 which 
might be responsible for the regularity of the outbursts. PDM was 
selected as the most appropriate technique owing to the irregularity of the modulation.
The signal-to-noise ratio appears good (the scatter of the open circles in Figure~\ref{fig:59amp} 
is small compared to the amplitude of the putative orbital modulation), and 59 second pulsations 
were detectable down to very low flux levels because the source was in the center of the PCA 
field of view. The resulting 
period is 123 $\pm$ 1 days, the folded lightcurve is shown in Figure~\ref{fig:59orbitnobins} 
where the zero point is MJD 50841, obtained by placing the peak at $\phi = 0$.

\subsection{AXJ0049-729 (74s)} 
AXJ0049-729 is one of the 3 pulsars discovered by \rxte in November 1997 \citep{IAU6803}.  The
low amplitude of the 74 s 
pulsations made it the weakest source in the discovery observation and slews were unable to
constrain its position accurately. An \ASCA observation on 1997 November 13 detected pulsations at
74.68(2)s \citep{Yokogawa99} and determined that AX J0049-729 lies in the error circle of the
\rosat source RX J0049.1-7250 \citep{KP98}. According to \citet{Stevens99} there is a single Be
star within the 13" error radius of the \rosat position, therefore this star is probably the
optical counterpart. Figure~\ref{fig:74amp} shows three separate outbursts occurred during the
\rxte monitoring program. Selecting only the 99\% significant detections (filled symbols in 
Figure~\ref{fig:74amp}) STA indicates a
candidate outburst-recurrence (orbital) period of 642$\pm$59 days. The flux measurements were 
folded at the 642 day period, and appear to be tightly constrained within a range of 0.3 in phase 
(Figure~\ref{fig:74orbit}).
Spectral parameters were obtained at the peak of the brightest recorded outburst on MJD 52078 and
the pulse profile shown in Figure~\ref{fig:74pro} is also from this observation. The spectrum was
well fit by the classic Be/X-ray binary model, parameters are given in Table~\ref{tab:catalogue},
the cutoff energy was 16.2$\pm$1.2 keV and a prominent iron K emission line was present.

\subsection{XTE J0052-725 (82s)}
Although pulsations at 82.4 s were regularly detected, the source was 
always extremely faint and the close similarity of its period to the harmonic of the 169s 
pulsar XTE J0054-720 made positive detection difficult. 
A brighter outburst was eventually observed in February 2002 (after the data presented in this 
paper). During the bright outburst slews were performed in order to localize the pulsar's 
position \citep{IAUC7932}. This position was then used retrospectively
to generate the lightcurve shown in Figure~\ref{fig:82amp}.

\subsection{AX J0051-722 (91s)} 
Pulsations with a period of 91.1 s were discovered in
November 1997 \citep{IAU6803} and were detected regularly throughout 1997 to 1999 as shown in 
Figure~\ref{fig:91amp} until a change in 
monitoring position shifted AX J0051-722 out to the edge of the field of view (see
Table~\ref{tab:positions}). \citet{IAU6858} reported that the source re-brightened on 1998 March
25, having faded since its initial discovery, suggesting an orbital period around 110
days. PDM analysis was performed on this subset of the data giving a period of 115 days, there is a
secondary minimum at 123 days and this may well be due to some degree of cross-contamination
from the nearby 59s pulsar XTE J0055-724 which has this orbital period, although the pulse periods
are not harmonics of each other. The folded lightcurve is shown in Figure~\ref{fig:91fold}. 
The 2-10 keV pulse profile in Figure~\ref{fig:91pro} is noisy due to the 
off-axis angle of the source, and shows a broad peak, roughly twice the duration of the 
pulse-minimum.

\subsection{XTE SMC95 (95s)}
\label{sect:95}
The 95 second pulsar was discovered on 1999 March 11 during the \rxte monitoring project.
An analysis of the initial discovery observations has been published by \citet{Laycock02a}. 
For reference purposes the source is provisionally designated \emph{SMC95}.
According to the additional data presented here (see Figure \ref{fig:95amp}), two separate 
outbursts were observed at $>$99\% significance in position 1. The lower panel of 
Figure~\ref{fig:95amp} shows that 2 or 3 more groups of observations reached greater than 90\% 
significance and these groupings are spaced at intervals roughly equal to the interval between 
the two strong outbursts. 
Since the position of the source is poorly constrained \citep{Laycock02a}, 95 s pulsations were
searched for in all observations and at all positions. The data obtained in positions 1a, 1b and 
1c had to be excluded due to the very bright 91 second pulsar AX J0051-722. The side-lobes of 
this pulsar contained so much power at 95 seconds that accurate estimates of \emph{SMC95} were 
impossible. A candidate orbital period was estimated with STA (Section~\ref{sect:sta}), giving 
280$\pm$8 days, for this period, we estimate the epoch of maximum flux T$_{0}$ = MJD 51248. 

\subsection{AX J0057.4-7325 (101s)}
This source appears in the long term lightcurve during 6 observations, only one of which attains 
a high detection significance. This is a result of the fact that AX J0057.4-7325 lies far from 
the center of the PCA field of view at all of the pointing positions. The strongest detection 
was during the deep observation described in section~\ref{sect:deep}. No spectrum was extracted 
for that observation because several other pulsars were active, and mostly at higher flux levels.
No orbital period could be estimated owing to the limited number of observations.

\subsection{XTE J0054-720 (169s)}
This source was discovered on 1998 December 17 with \rxte \citep{IAU6814}.
Except for its first detected outburst it was a faint source as seen in Figure~\ref{fig:169amp}.
Timing analysis of the lightcurve did not provide conclusive evidence of an orbital period. 
PDM gave weak minima at 112 and 224 days using the full dataset. Selecting only the 99\% 
significance detections (17 points) and applying the simplified timing analysis produced 
minimum scatter in the detection dates at 53$\pm$11 and 201$\pm$41 days, the former is seemingly
ruled out because it is shorter than the well observed outburst at the beginning of the dataset.
This outburst has an apparent duration of about 100 days and is brighter than the other outbursts. 
and there is a pattern of 3 narrow minima in its significance plot. The large outburst 
could therefore be anomalous, extending far beyond periastron. The true period could then be 
close to the lower figure. The result of folding the lightcurve at a period of 224 days is shown 
in Figure~\ref{fig:169fold}.
 

\subsection{AX J0051.6-7311 (172s)}
Pulsations at 172.4 s were reported by \citet{Torii00a} in an \ASCA observation in April 2000.
\rxte monitoring has revealed AX J0051.6-7311 to be one of the most frequently active pulsars in 
the SMC. The source has been identified with the \rosat source RX J0051.9-7311 and a Be star found 
in the error circle by \citet{Cowley97}.

The pulsed flux for AX J0051.6-7311 was generally $<$0.5\fluxpcu, placing it at the low luminosity
end of the SMC pulsar population. Despite many detections over 
a long timebase, the orbital period remains elusive. A number of 
approaches were tried in an effort to determine an approximate value for the recurrence time 
between outbursts. For this purpose we used only the position 5 data (after MJD 516000), with 
the pulsar well placed in the PCA field of view.
  
After removing all points with non-zero detection significance we were left with 52 possible 
measurements of the pulsed flux. PDM analysis of this lightcurve  
gave no conclusive period. Under the assumption that detections are more likely to be close to 
orbital phase zero, and seeing a series of at least 6 equally spaced peaks in Figure~\ref{fig:172amp}, 
the 99\% detections (20 points) were analyzed with STA, giving a period of 64$\pm$16 days. 
Selecting only the 6 evenly spaced outbursts (MJD 51686 - 52039), the recurrence 
period is 67$\pm$5 days, T$_0$ = MJD51694. It seems likely that in addition to strings of periodic 
outbursts this pulsar exhibits some outbursts that are not closely correlated with orbital phase.

\subsection{RX J0050.8-7316 (323s)}
323s pulsations were first detected by \ASCA on 1997 November 
13 \citep{Yokogawa98a} at a position coincident with the \rosat source RX J0050.8-7
316. The optical 
counterpart is thought to be a Be star identified by
\citet{Cowley97}. \citet{CoeOrosz} have shown that this star is in a binary system with a 
1.4 day period. RX J0050.8-7316 is a particularly interesting system as it has recently been 
proposed as a triple system \citep{trinary}.
   
Pulsations consistent with a 323 s period were detected frequently with \rxte throughout the survey
and the source was near the center of the field of view for the majority of the time (MJD 51555
-52333). The long pulse period and low luminosity result in fairly large uncertainties on both 
period and flux determinations for most observations, however there seem to be no other pulsars 
with periods that are likely to cause confusion in the 300s - 330s period range. 
Although AX J0103-722 has a pulse period variously reported as 343s or 348s it was not in the 
position 5 field of view and therefore cannot be present in the lightcurve 
(Figure~\ref{fig:323amp}) after MJD 51555.

The signal to noise level for the long term lightcurve was rather low, especially 
in the early observations obtained in position 1. During these observations the source was 
0.85$\degr$ off-axis and hence our sensitivity was poor. 
The 99\% significance detections obtained 
during 1998 - 2002 (23 black points in Figure~\ref{fig:323amp} after MJD 51200) were analyzed using 
STA, giving a period of 108$\pm$18 days. This procedure did not take direct account of the flux 
values and is based on the assumption that each detection corresponds to X-ray emission at or close 
to phase zero (outburst peak). For a faint source this seems a reasonable assumption considering 
our sensitivity limit is approximately 0.2 \fluxpcu for most observations. 

Having identified a tentative period, the earlier observations from 1997-1999 were included in the folded 
lightcurve shown in Figure~\ref{fig:323fold}, the 6 brightest points 
belong to the earlier position 1 data. 
Having identified a candidate orbital period we propose the peak of the best observed outburst 
be used as epoch of phase zero, this is MJD 51651.

\citet{Imanishi99} performed an analysis of archival data from \ascac, 
\rosat and \emph{Einstein} (total 18 observations), finding weak evidence for a periodicity 
of $~$185 days.

\subsection{AX J0103-722 (348s)}
This pulsar appeared in 4 observations in Figure~\ref{fig:348amp}, which could be used to estimate 
some kind of recurrence timescale. However the source was not near the center of the field of view 
in any position except position 2 (which was only observed 4 times). 
The \ASCA pulsar AX J0103-722 has been identified with a \rosat source lying in a supernova 
remnant with a Be optical companion. Detections have been found in data from \Einsteinc, \SAXc, 
\rosatc, and \Chandra all at luminosities of $\sim$10$^{36}$\ergps, the infrequent detections 
by \rxte are therefore explainable by the low luminosity of the source. In fact at 
10$^{36}$\ergps\ it would be near the sensitivity limit for most observations.

\subsection{RX J0101.5-7211 (455s)}
Pulsations consistent with the 455s pulsar RX J0101.5-7211 were detected at low flux levels.
The source was discovered by \rosat \citep{Haberl00}, noted to be highly variable and 
was identified as a Be/X-ray binary. Although Figure~\ref{fig:455amp} appears somewhat noisy 
there is no other known pulsar with a period or harmonics likely to cause confusion. 
We note the presence of approximately equally spaced points at $>$ 90\% significance which are 
suggestive of an outburst spacing of 200-300 days. Only 2 points reach our nominal detection 
threshold (99\%) and clearly further observations are required.

\subsection{AX J0049.4-7323 (755s)}
\label{sect:755}
This source has the longest pulse period so far seen in the SMC.
It was discovered with \ASCA on 2000 April 11
\citep{Ueno00b,Yokogawa00a} and the reported \ASCA pulse period was 755.5(6)s. \citet{Yokogawa02} 
described how, in addition to the discovery observation, the source was also detected by \ASCA on 
1997 November 13 and 1999 May 11 but with no detection of pulsations, presumed 
due to the short duration of these observations. The reported luminosities during the \ASCA detections
of AX J0049.4-7323 all appear to be around 5$\times$10$^{35}$ \ergps -well below the \rxte 
sensitivity threshold. According to \citet{Yokogawa02} a revised analysis of the \ASCA position 
associates AX J0049.4-7323 with the \rosat source RX J0049.5-7310.

Figure~\ref{fig:755amp} shows that \rxte saw AX J0049.4-7323 in two separate outbursts, centered 
around MJD 51800 and MJD 52200. The mean pulse period measured by \rxte was 751s. Based on the 
spacing of the outbursts and the scatter of points, a candidate orbital period was estimated, 
to be 396$\pm$5 days with STA (Section~\ref{sect:sta}). The zero point to place the observed outburst 
peak at $\phi$ = 0 is $T_{0}$ = MJD 51800.  
Obviously a ``period'' derived from two outbursts would be considered provisional until 
future observations confirm it. Such confirmation for this outburst interval
being the orbital period has indeed now come from optical measurements by \citet{Schmidtke04} who found 
optical outbursts from this system every $\sim$394 days.

The second outburst is remarkable in its apparent brightness, the pulsed flux was 
approximately 3.2 \fluxpcu on 2001 October 11, making it one of the most luminous pulsars 
in the SMC. The pulse profile for that observation is shown in Figure~\ref{fig:755pro}. 
A spectrum was also extracted for this observation (full parameters in 
Table~\ref{tab:catalogue}) implying an unabsorbed 
$L_{X}^{2-10}$= 1.7$\times$10$^{37}$\ergps, a high energy cutoff was required at 12.16$\pm$1 keV 
and an iron K line was also present.

Archival data from \rosat and \Einstein were analyzed by \citet{Yokogawa00a}, demonstrating 
that AX J0049.4-7323 has been active at the $\leq$5$\times$10$^{35}$ \ergps\ level for over 20 years. 
Activity at this level falls below our detection threshold and we are
apparently only detecting relatively infrequent large outbursts.
It is noted that the pulse profile obtained for the 3-10 keV 
range (Figure~\ref{fig:755pro}) bears no resemblance to the \ASCA pulse profile 
suggesting that the pulse profile is very luminosity-dependent.

\section{A Very Deep Observation}
\label{sect:deep}
One very deep observation was made at Position 4 (see Table~\ref{tab:positions}). Beginning on 
MJD 51801.2 a total of 106.5 ksec of good time, spanning nearly 3 days was obtained. 
The standard analysis as described above revealed the presence of 7 pulsars, of which 2 were new 
discoveries. The dataset was included in the long-term lightcurves presented above. The periodogram 
is shown in Figure~\ref{fig:longls} with all of the 99\% significant peaks labeled and identified.    
Because this observation spanned a very long time interval, the periodogram was 
calculated out to 100 ksec in order to search for hitherto undiscovered long-period pulsars. 
A number of features are evident at low frequencies and most of these are attributed to 
systematic effects related to the \rxte orbital period and residual diurnal background variations.
Pulsars detected are (from left in Figure~\ref{fig:longls}) RX J0049.7-7323 and several harmonics, 
RX J0051.9-7311, AX J0057.4-7325, XTE 51s (detected P/2 harmonic), XTE 16.6s, RX J0052.1-7319, 2E 
0050.1-7247.


\section{Known SMC Pulsars Not Detected With \rxte}

A table of known SMC X-ray pulsars which were not detected in any of the monitoring observations 
despite being in the \rxte field of view at various times is provided, see 
Figure~\ref{tab:nondetect}.
In each case a single upper limit on the flux is quoted, 
based on the observation judged to be the most sensitive 
in terms of position, duration, and number of detectors functioning. 

\section{General Properties of the SMC HMXB Population.}

\subsection{Pulse Periods}
We compare the pulse period distributions for X-ray pulsars in the 
SMC, LMC and the Galaxy. Figure~\ref{fig:dist_smclmcgal}
shows the 3 populations binned with 2 bins per decade in period. A comparable  
number of pulsars are known in the SMC (30) and Galaxy (54), and a lesser number in the LMC
\citep{Liu00,Sakano00,Marshall00,Intzand01,Bamba01}.

The pulse periods in Figure~\ref{fig:dist_smclmcgal} mainly occupy the 10 - few 100 s range,
with the Galactic population skewed toward longer periods. The known range of pulse periods 
stretches from below 0.1s to over 1000s. 
We compared the Galactic and SMC populations with the Kolmogorov-Smirnov test.
The K-S statistic was 0.3222, giving a probability of 0.028 that the two samples are drawn from 
the same population. Some caution is required in interpreting this result because the conditions 
under which the two samples were obtained were far from homogeneous. If one were to try to 
characterize typical observing conditions for the SMC pulsars the important factors would be low 
galactic extinction, and constant distance ($\sim$65kpc). For the Galactic pulsars the conditions 
are varying extinction across as much as several decades in $N_H$ and a wide range of distances 
which are uncertain by factors of 2 or more.

If systematic differences exist between the pulse period distributions in the two galaxies, 
physical causes must be disentangled from selection effects. 
Our results presented here tend to support the link between X-ray luminosity and pulse period. 
\citet{SWR86} pointed out that the  maximum observed luminosity for binary
X-ray pulsars is anti-correlated with $P_{pulse}$, although the results presented here suggest 
that the effect is weak for periods longer than a few seconds. Luminosity 
selection bias (against fainter long period pulsars) in our survey seems unlikely to
seriously affect results. A more 
serious bias against long period pulsars is observation length: 
3 ksec is only long enough to detect pulsars up to periods of about 750 seconds. 
At very short pulse periods, there is unlikely to be any luminosity bias because these sources
are bright (typically several 10$^{37}$\ergps). The important factor for 
short period pulsars is density of observing coverage because these systems rarely 
go into outburst. It seems unlikely that any other pulsars similar to SMC X-2 and XTE 
J0052-723 could have been missed because the typical duration of these outbursts is  
several weeks. With observations on a weekly basis at a sensitivity $\sim$10$^{36}$\ergps\ any 
such pulsars would have been detected if they lie within the field of view of the PCA. 
XTE J0052-723 was easily detected at a collimator response of just 0.2. A look at the Corbet 
diagram for Galactic and SMC pulsars (Figure~\ref{fig:newcorbet}) reveals that the cause of the 
discrepancy in pulse period distributions is in fact the predominance of Be/X-ray binaries in the SMC. 

Our luminosity sensitivity of $\sim$10$^{36}$\ergps\ may be converted into an
approximate pulse period detection threshold. \cite{SWR86} found an inverse correlation between pulse period and
maximum X-ray luminosity. From Figure 2 of \citet{SWR86} our luminosity sensitivity implies that we would 
be able to detect SMC pulsars with periods shorter than $\sim$1000 seconds if SMC pulsars follow the same 
relationship. We note that \citet{Majid04} do find a comparable relationship between maximum 
luminosity and pulse period for the SMC and Galactic sources.

For Galactic HMXBs, a significant fraction of long period
pulsars are accreting from the winds of supergiant companions.
These systems, although variable, are persistent which
makes them much easier to detect than transient Be
star systems. In spite of this, none of the SMC
pulsars under discussion here has the characteristics
of a supergiant wind-fed binary. Therefore, one
contribution to the difference between the Galactic
and SMC pulse period distributions is the apparent lack of
supergiant wind-accretion pulsars in the SMC.

\subsection{Orbital Periods}
Transient Be star X-ray pulsars exhibit two types of outbursts
\citep[e.g.][]{Bildsten97}.
Type I outbursts recur, when a system is in an active state,
on the orbital period of the system. Activity is confined to
limited orbital phases around the time of periastron
passage. This activity pattern is the most common type
of outburst. In addition, Type II (giant) outbursts may also occur
where the X-ray luminosity is much larger and is not
modulated on the orbital period.
Our observations of SMC pulsars (over a timescale very much longer than the expected orbital periods) shows 
detections of repeat outbursts in 10 sources. Under the assumption that
many of these are the more common Type I outbursts, these
should allow the estimation of orbital periods for several
systems.

Candidate orbital periods and their uncertainties are listed in Table~\ref{tab:catalogue}. 
Under the widely discussed ``standard'' model of Be/X-ray binaries \citep[e.g.][]{Negueruela98}, 
it is expected that Type I outbursts will occur periodically as the neutron star's eccentric orbit 
intercepts the Be star's circumstellar disk, while type II outbursts are triggered by large-scale 
enhancements in the disk. 
Orbital periods for the SMC pulsars and for all known Galactic HMXB systems are shown in 
Figure~\ref{fig:newcorbet}, an updated version of the $P_{spin}/P_{orbit}$ diagram 
\citep{Corbet86}. Uncertainties in $P_{orbit}$ are indicated, uncertainties in $P_{pulse}$ are 
smaller than the plot symbols. 

The $P_{spin}/P_{orbit}$ diagram demonstrates that X-ray pulsars with massive 
companions display three different correlations between their pulse and orbital periods. 
These correlations have been successfully explained by \citet{Corbet86} in terms of the specific 
mode of mass 
transfer occurring in the binary. Different regions of the diagram are populated by systems 
undergoing steady wind-fed accretion, Roche lobe overflow, and transient (often periodic) accretion 
as is the case for Be systems. A powerful feature of the Corbet diagram is that the optical 
counterpart to an HMXB can be predicted to a high degree of confidence if only the timing 
parameters are known.  
Optical counterparts for around half of the currently known SMC pulsars have been 
identified and classified. The similarity of the X-ray properties of virtually all SMC pulsars 
(excepting SMC X-1) suggest the unidentified sources are also Be/X-ray binaries. 
In summary this evidence comprises: (1) the transient nature of the X-ray emission, 
(2) the typical luminosities and (3) the spectral parameters. The picture established from X-ray 
observations is supported by the optical identifications all of which are Be, excluding SMC X-1 
which seems to be the lone example of its class. 
There are 32 HMXBs in the Galaxy for which both orbital \emph{and} pulse 
periods have been reported, and 2 in the LMC \citep{Liu00,Intzand01b,Delgado01,Intzand01}. 
In Figure~\ref{fig:newcorbet} color coding used to signify the 3 types 
of high mass accreting systems. Square, triangle, and star symbols indicate Galactic, LMC and SMC 
sources respectively. 

All 8 of the newly determined candidate orbital periods for SMC pulsars lie in the upper half of 
the Be distribution as defined by the Galactic systems. Despite the pulse 
period distribution being weighted (although weakly) in the opposite direction. 
There is no characteristic of our \rxte observing strategy that would be expected to cause a 
bias toward finding pulsars with such long orbital periods, although the weekly sampling 
makes orbital periods below about 14 days difficult to measure. The apparent lack of short orbital 
periods for the SMC population is instead due to the small number of pulsars detected 
with pulse periods below $\sim$10s. 
For periods approaching 1 second, the minimum accretion-driven luminosity for a 10$^{12}$ 
G neutron star is close to the Eddington limit. So fast spinning pulsars are expected to 
spend the majority of the time in the centrifugal inhibition regime (if they are in Be systems). 
Thus unless a large dense Be star disk is continuously present, regular 
type I outbursts cannot occur. For such systems the orbital period can only be 
determined from pulse timing analysis or monitoring of the optical counterpart.    

\section{Spectral Parameters}
HMXB spectra are in general characterized by a single power 
law, modified by absorption at low energy and an exponential cut-off at high energy \citep[e.g.][]{White83}. 
In systems where a cyclotron
scattering resonance feature has been seen, which
provides a measurement of the neutron star magnetic field
strength, the spectral cutoff energy is found to be correlated
with cyclotron energy \citep[e.g.][]{makishima92,coburn02}. Measurements of spectral cutoffs thus
provide at least estimates of the magnetic field strength. 
A fluorescence line is also often seen at 6.4 keV 
corresponding to the iron K line from relatively cool material \citep{Nagase89}.  

For each bright pulsar observed during the SMC monitoring, a spectral fit was performed for at 
least one detection. Spectra were only extracted when no other pulsar was active, to avoid 
cross-contamination. The resulting spectral parameters given in 
Table~\ref{tab:catalogue} show values consistent with those
exhibited by Galactic sources \citep{White83}. 

Iron lines were detected in most SMC pulsars: typical line widths were 
around 0.5 keV and the line strengths ($\sigma$) were 
correlated with overall luminosity. One exception to the above was XTE J0055-724 
(period 59 s) which was X-ray bright but showed no trace of an iron K line. Three sources stand 
out as having peculiar 
features at the high energy end of their PCA spectra. The 74.7s, 172.4s and 323s pulsars all 
show very steep cutoffs although their spectral indices below the 
cut-off are not unusual. In addition these fits are not good representations of the observed 
spectra above the cutoff and the residuals show the possible presence of either a deep absorption 
feature around 20 keV or an emission feature at slightly lower energy. The two most notable 
of these are also the most frequently active sources in the SMC. 
The 172.4s and 323s pulsars were found to exhibit at least 5 and 7 outburst respectively. 
The pulsed-lightcurves presented in Section~\ref{sect:lcurves} show 
that the situation with these two sources is complex and there may be faint pulsations 
emitted at scattered times not clearly correlated with orbital phase. This interpretation 
of the pulsed lightcurves could imply that the pulsars are not being centrifugally inhibited
at low accretion rates. Such a situation would be consistent with a lower value 
for the magnetic field which would be expected to reveal itself as unusually low 
energy cyclotron lines in the X-ray spectrum.

\section{The Spatial Distribution of HMXB in the SMC}
HMXB provide tracers of recent star formation activity because they are short lived, 
make an appearance very soon in the evolution of a 
star forming region, and are of course highly conspicuous. It has been noted that many X-ray 
pulsars discovered in the Galaxy are concentrated in the nearby ``5 kpc arm'',
a region identified by radio, IR, and CO molecular line emissions \citep{Hayakawa77}. 
Six pulsars discovered with the \emph{Ginga} satellite are known in this region and all are 
transient sources having the X-ray characteristics of Be systems. Spiral arms in galaxies are 
well known to harbor star forming regions and in fact such regions account for the majority of 
star formation within galaxies.
The work of \citet{mgm99} proposes that the ratio of Be to normal B type stars is higher in 
the SMC than in the Galaxy, although this was based on observations of certain SMC clusters 
which are probably not representative of the SMC as a whole. 
A promising line of inquiry would be to carry out a similar survey concentrating on comparing 
the regions harboring concentrations of HMXB.
In order to identify the regions to observe, a large number of HMXB are needed for significant 
clustering to become apparent. Be stars are themselves relatively 
conspicuous objects from their photometric colors, and their place as the most common 
optical counterpart in HMXB undoubtedly has strong implications for the preferred channel for 
massive binary star evolution. 

Recent optical surveys of the SMC \citep{Zaritsky00, Cioni00, Maragoudaki01} show 
that the young stellar population is concentrated into distinct structures. 30 X-ray 
pulsars have been identified in the SMC, the majority with positions to the \amin\ level. 
These positions are plotted in Figure~\ref{fig:spatial} and can be compared with the distributions 
of other SMC constituents. Neutral hydrogen is a natural galactic constituent to compare with 
the HMXB distribution because large concentrations of hydrogen are 
a necessary precursor to star formation. Figure~\ref{fig:spatial} panel 3 shows the HI density 
distribution as mapped by \citet{S99}, superimposed are the positions of all the SMC pulsar with 
accurately known positions. 
\citet{Maragoudaki01} produced isodensity contour maps of the SMC showing the spatial 
distributions of stars of different ages. Perhaps the most interesting panel from the point of 
view of HMXBs is the distribution of stars whose age is comparable to the evolutionary timescale 
for HMXB formation. The isochrone map for stars aged 8 - 12.2 My is reproduced here in 
Figure~\ref{fig:spatial} panel 2. A catalog of emission-line stars in the SMC has 
been compiled by \citet{MA93} based on an objective-prism survey. The catalog was searched for 
all objects with a positively detected H$\alpha$ emission-line, in a 5 degree wide field centered 
on the SMC, and the results are plotted in Figure~\ref{fig:spatial} panel 1. The final population 
distribution plotted here is the X-ray source population as seen by \ROSAT with the PSPC and HRI. 
The PSPC catalog is larger due to greater sensitivity however the HRI catalog gives the 
possibility to restrict the search to point-sources only. This distinction enables at least an 
approximate restriction to a sample of HMXB. There is some observing bias surrounding the \ROSAT 
distribution because the 5$\degr \times$5$\degr$ field  was not uniformly observed by either 
\ROSAT instrument, it appears that the main body of the SMC was evenly covered while the two 
apparent clusters around (00$^h$ 40$^m$, -72\degrees) and (01$^h$ 20$^m$, -75\degrees) are due to two particular pointings. The 
cluster of sources in the ``wing'' is real.
 
Looking at Figure~\ref{fig:spatial} it is apparent that there is a close correlation between 
the spatial distributions of the 5 populations. The SMC bar and wing are clearly marked 
out by the clustering of X-ray sources, emission-line stars, young stars and HI. 

\section{Summary}
The SMC is an intriguing galaxy because of its surprisingly high abundance of X-ray pulsars. 
The X-ray binary population of the SMC shows important differences from the LMC and the Galaxy. 
The SMC appears to have been chemically isolated from the Galaxy and has been sculpted by  
gravitational and hydrodynamic forces originating in tidal interactions with the LMC and the Galaxy.
As such the SMC provides an ideal environment in which to study the effects of dynamic encounters 
on star formation and the evolution of the stellar population. A number of recent works have 
revealed patterns in the spatial distributions of different aged stellar populations, and this 
evidence can be used to constrain various models of stellar evolution and structure formation. 
The large population of pulsars provides both a probe of the star formation and an ideal sample 
to investigate the general properties of HMXB. 


\appendix
\section{Catalog of Sources Detected}

Parameters are summarized here for each pulsar detected during the project. 
For every known pulsar in the SMC,  name(s), position and optical counterpart can be found 
in Table~\ref{tab:pulsars}. This table should be used to identify the pulsars for which 
detailed measurements were made during the \rxte monitoring project. Parameters determined in 
this work are summarized in Table~\ref{tab:catalogue} and concern the subset of 18 pulsars which 
were observed under particularly good signal to noise conditions. For these pulsars the 
minimum and maximum pulsed fluxes, spectral parameters, and luminosities are presented. 

A complete description of the columns in 
Table~\ref{tab:catalogue} is given below:

\small  
\noindent
(1) \emph{Pulse Period.} All of the pulsars exhibited pulse period variations, this column lists
a characteristic value. 
\newline
(2) \emph{Orbital period.} Measured in days by the methods described in the relevant sections of
 this paper.
\newline 
(3) \emph{$T_0$ Epoch of Phase 0.} Orbital period zero point, where phase zero is taken to correspond with
peak X-ray emission.
\newline
(4) \emph{Minimum pulsed flux.} Units of
\fluxpcu. Lowest pulsed flux measured for the source at better than 
99\% significance. Determined from the power spectrum as described in Section~\ref{sect:lcurves}. 
\newline
(5) \emph{Maximum pulsed flux.} Units of \fluxpcu. 
\newline
(6)-(16) \emph{Characteristic Spectral parameters.} Taken from selected observations based on 
brightness and lack of interfering sources. Fluxes are the absorbed values in 10$^{-11}$\fluxerg. 
With the exception of (L$_x$) the values have not been corrected for collimator response or distance to the SMC.   
\newline
(9) \emph{Absorption column density} for neutral hydrogen in units of 10$^{22}$cm$^{-2}$ 
\newline
(14) Reduced $\chi^{2}$ for the fit. Fits were performed over the range 3 - 20 keV. 
\newline
(15) \emph{Collimator response} for the pulsar in the observation from which the spectral fit was 
performed. 
Numbers in parentheses are the pointing positions. Both parameters refer to Table~\ref{tab:positions}. 
\newline
(16) \emph{Luminosity} in units of 10$^{37}$ erg s$^{-1}$ The unabsorbed source luminosity assuming 
a distance of 65kpc after correction for collimator response. The 2-10 keV luminosity is 
given to facilitate comparison with results from other missions. \newline

\normalsize


\clearpage

\clearpage

\clearpage

\clearpage

\begin{figure}
\includegraphics[width=10cm,angle=-90]{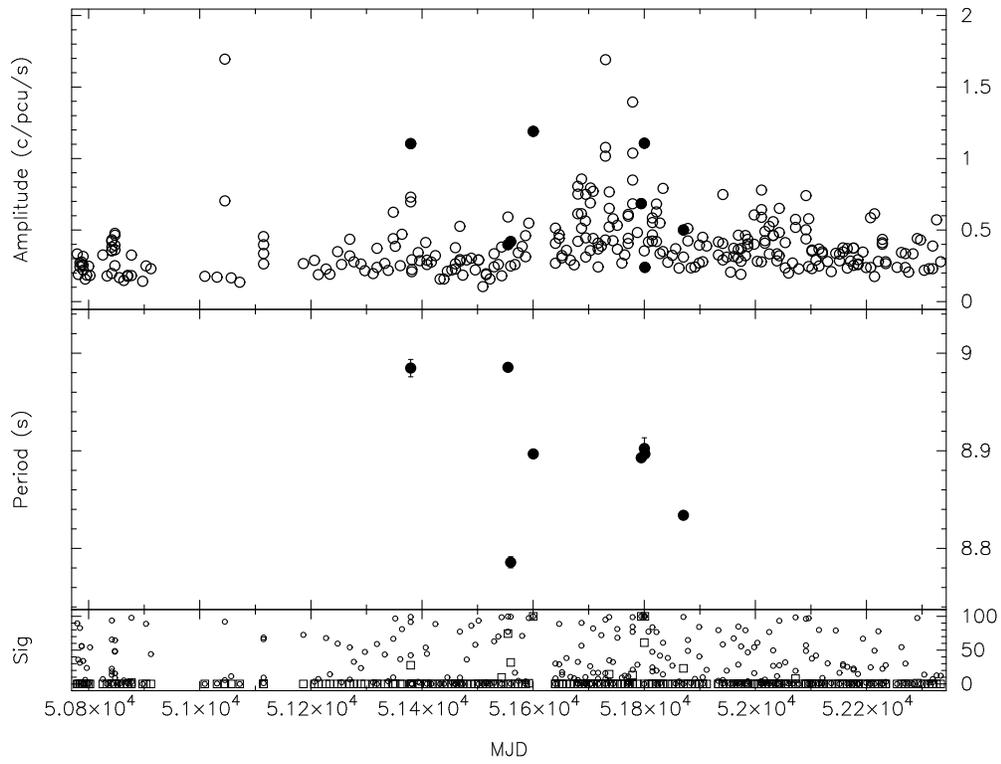}
\caption{2E 0050.1-7247, XTE 8.88, pulsed flux and period history.}
\label{fig:8.88amp}
\end{figure}\clearpage

\begin{figure}
\includegraphics[width=10cm,angle=-90]{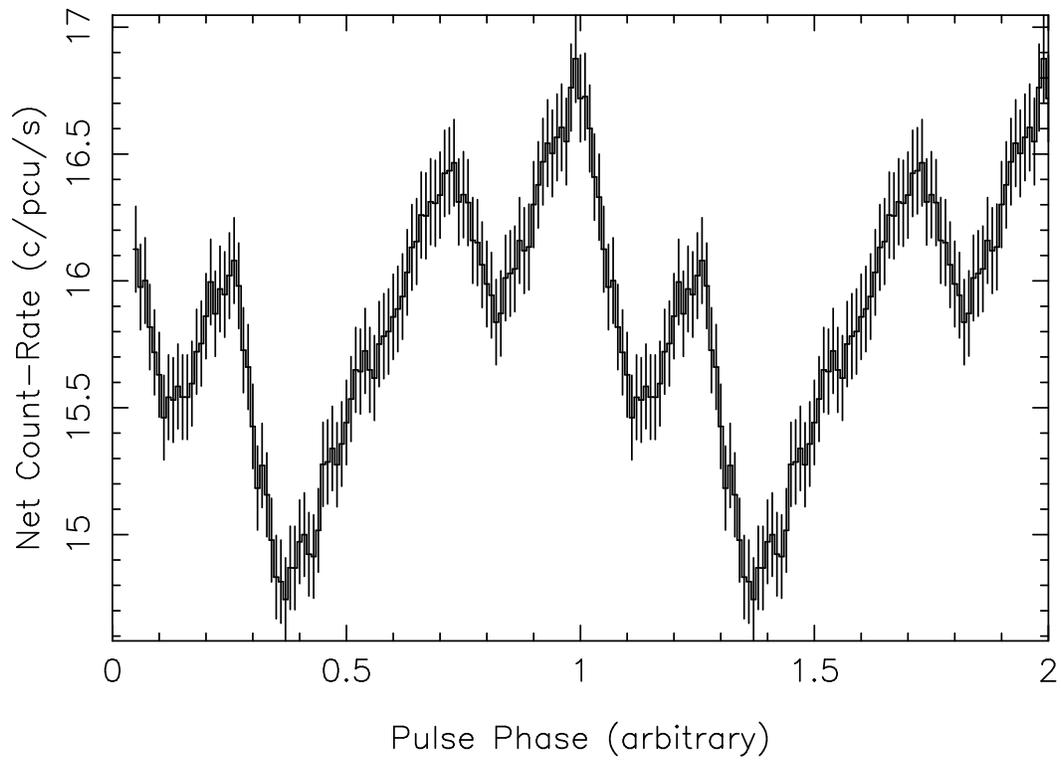}
\caption{Pulse profile for the 8.88s pulsar 2E 0050.1-7247.}                            
\label{fig:8.88pro}
\end{figure}\clearpage

\begin{figure}
\includegraphics[width=10cm,angle=-90]{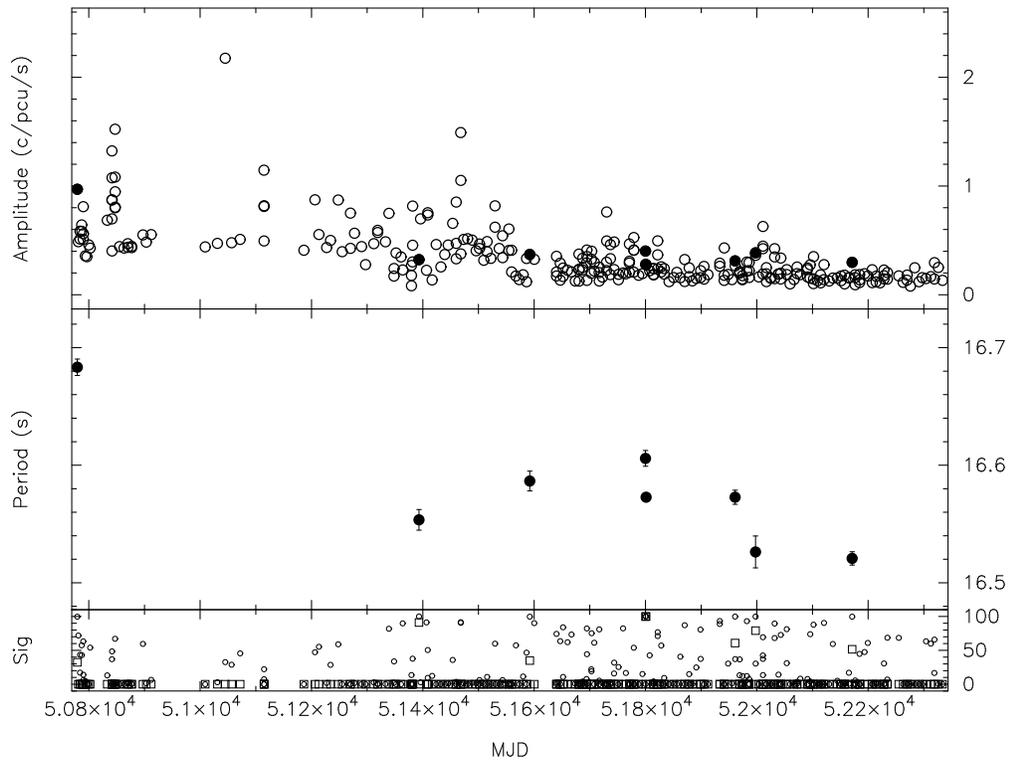}
\caption{XTE 16.6s, (may be AX J0058-720) pulsed flux and period history.}
\label{fig:16amp}
\end{figure}\clearpage

\begin{figure}
\includegraphics[width=10cm,angle=-90]{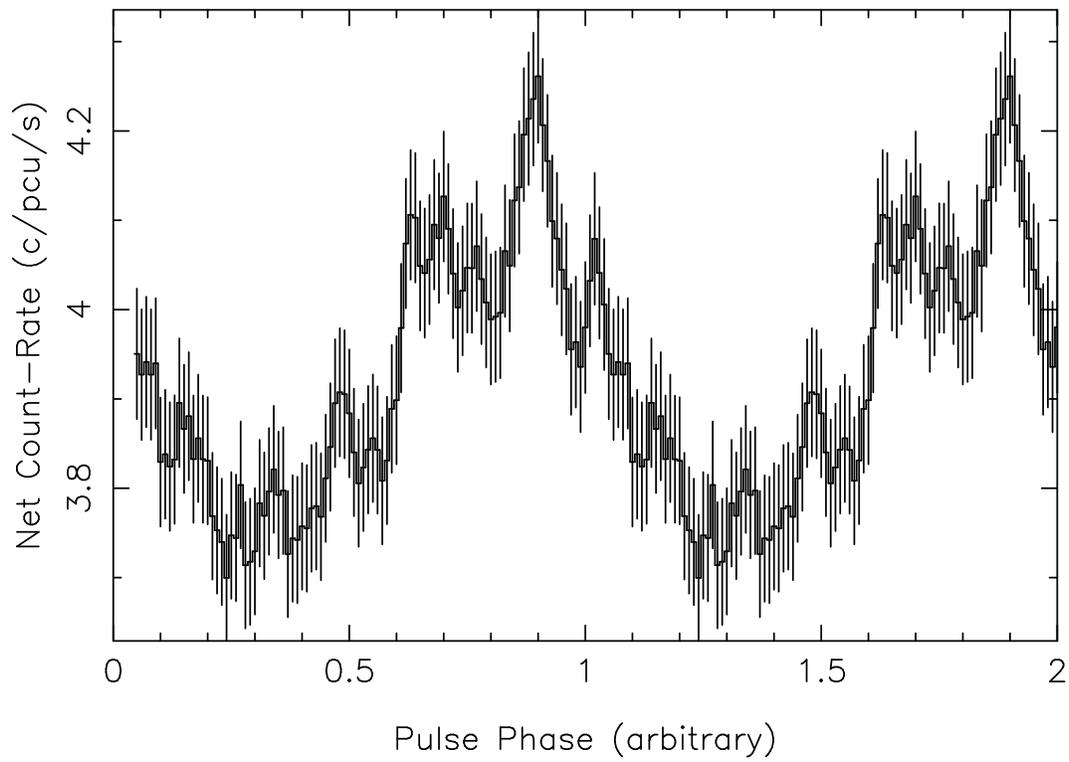}
\caption{Pulse profile for the XTE 16.6 pulsar (may be AX J0058-720), flux is in counts per PCU per second.}
\label{fig:16pro}
\end{figure}\clearpage

\begin{figure}
\includegraphics[width=10cm,angle=-90]{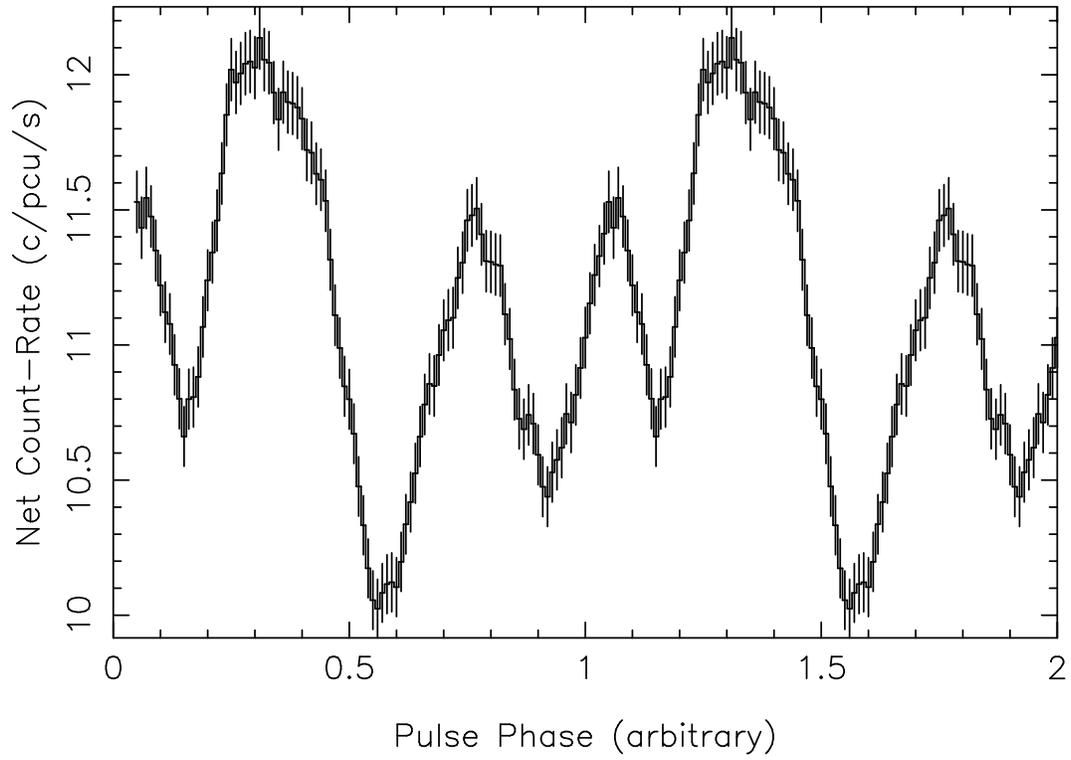}
\caption{Pulse profile for the 31s pulsar XTE J0111.2-7317}
\label{fig:31pro}
\end{figure}\clearpage

\begin{figure}
\includegraphics[width=10cm,angle=-90]{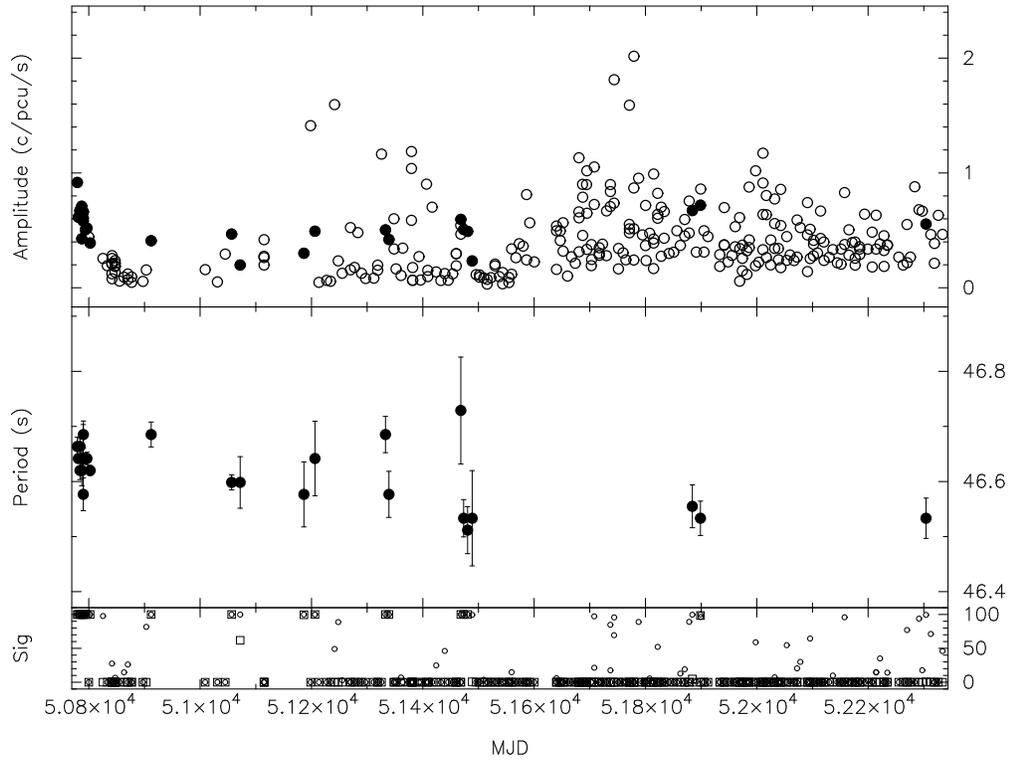}
\caption{XTE 46.6 (1WGA J0053.8-7226) pulsed flux and period history.}
\label{fig:46.6amp}
\end{figure}\clearpage

\begin{figure}
\includegraphics[width=10cm,angle=-90]{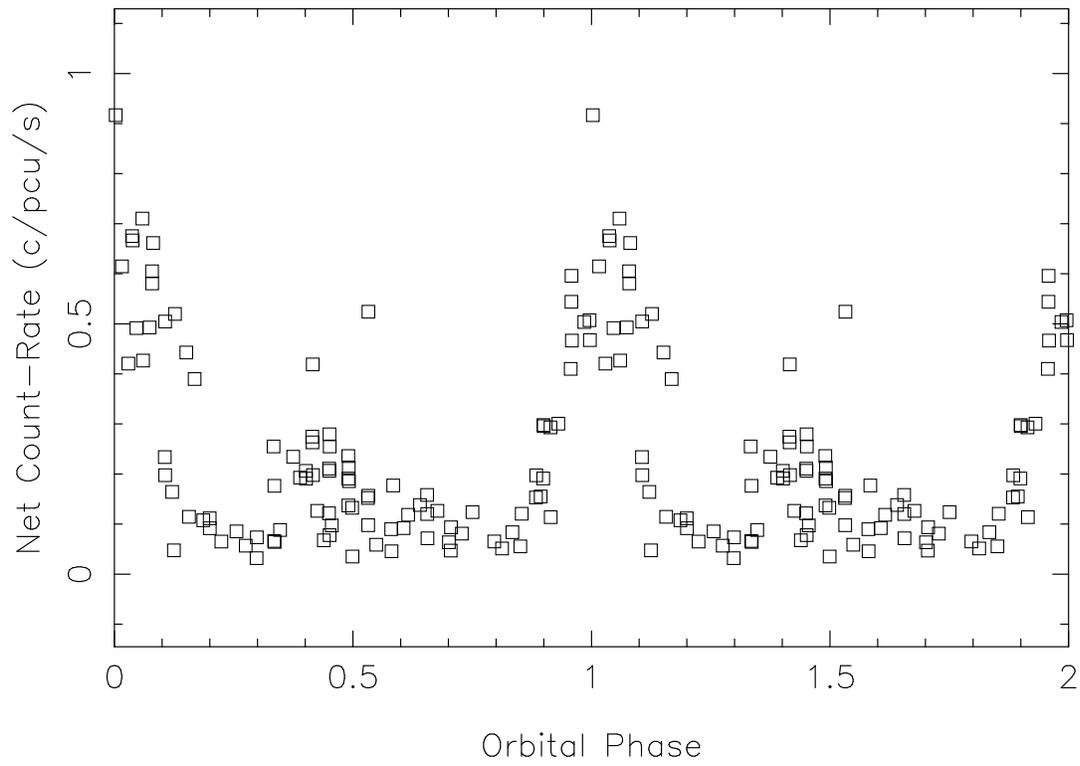}
\caption{XTE 46.6. Folded pulsed flux lightcurve at a period of 138 days. Note the  
feature at a phase of 0.5 which is possibly also visible in the unfolded lightcurve.}
\label{fig:46fold}
\end{figure}\clearpage

\begin{figure}
\includegraphics[width=10cm,angle=-90]{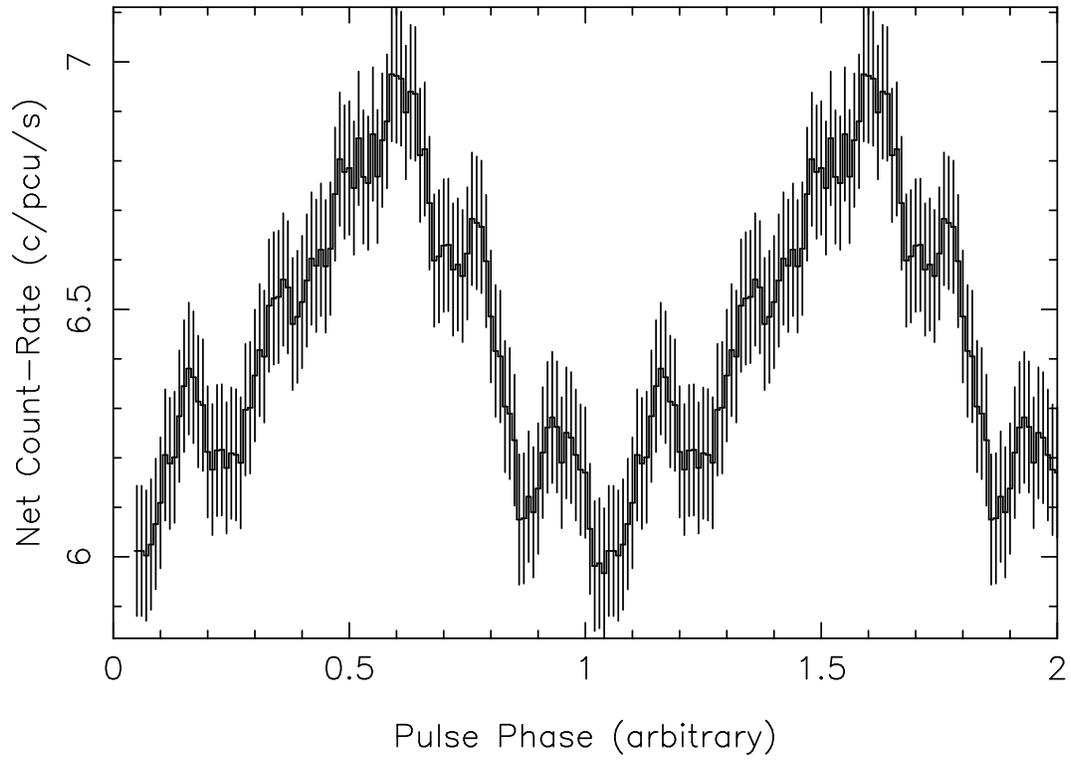}
\caption{Pulse profile (3-10 keV) for the 46.6 second pulsar 1WGA J0053.8-7226.}
\label{fig:46pro}
\end{figure}\clearpage

\begin{figure}
\includegraphics[width=10cm,angle=-90]{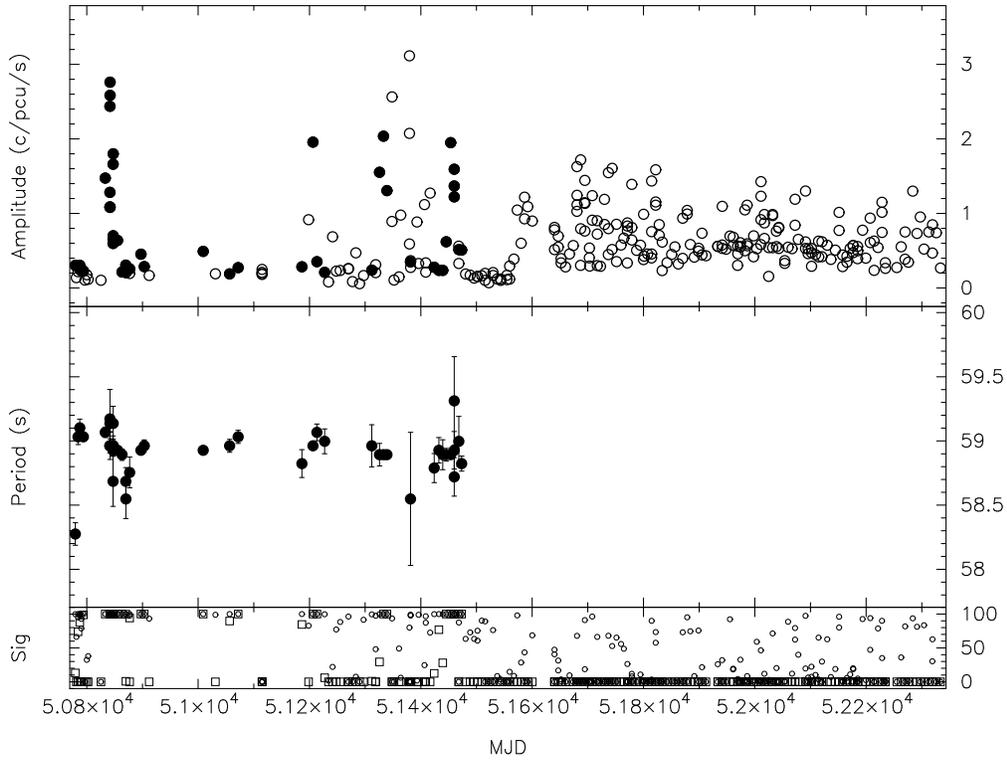}
\caption{XTE J0055-724. Pulsed flux and period history.}
\label{fig:59amp}
\end{figure}\clearpage

\begin{figure}
\includegraphics[width=10cm,angle=-90]{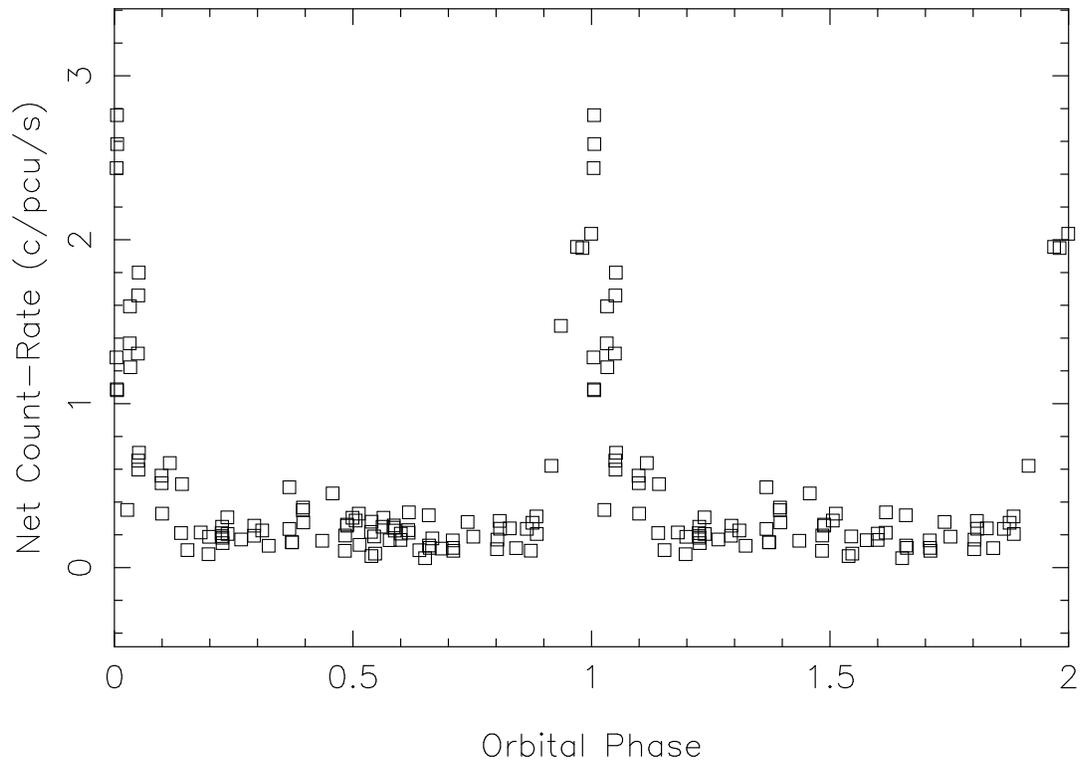}
\caption{XTE J0055-724 pulsed flux lightcurve folded at 123 day orbital period.}
\label{fig:59orbitnobins}
\end{figure}\clearpage

\begin{figure}
\centering
\includegraphics[width=10cm,angle=-90]{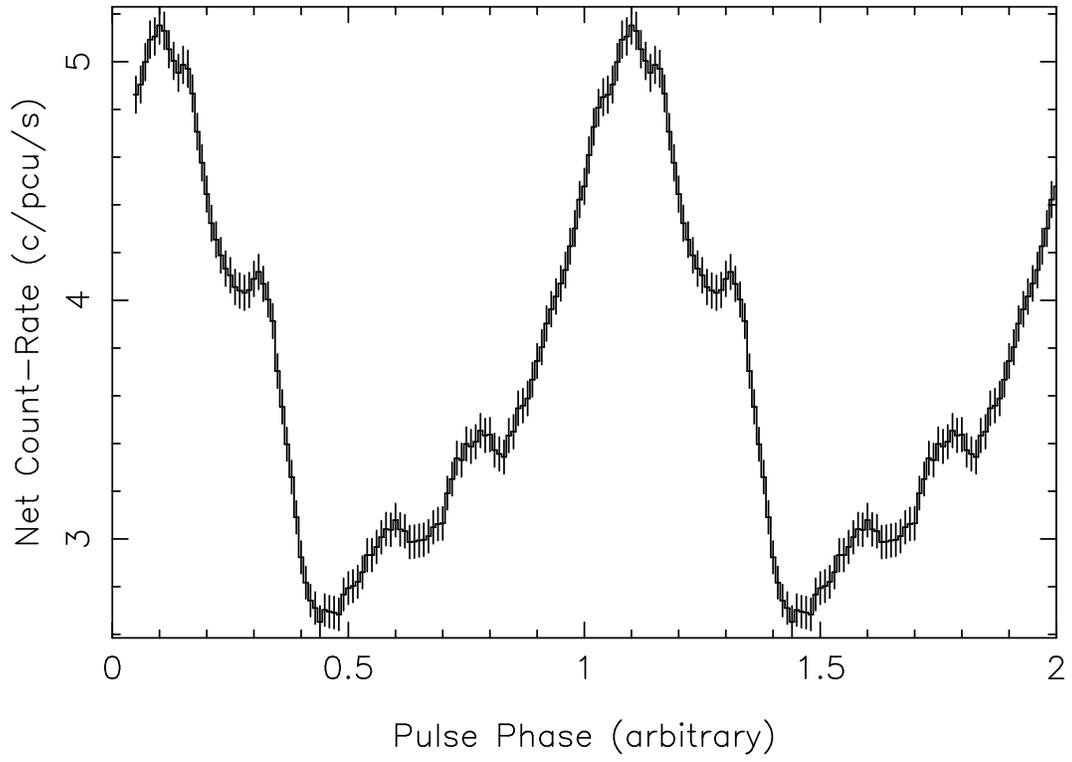}
\caption{Pulse profile of the 59 s pulsar XTE J0055-724}
\label{fig:59pro}
\end{figure}\clearpage

\begin{figure}
\includegraphics[width=10cm,angle=-90]{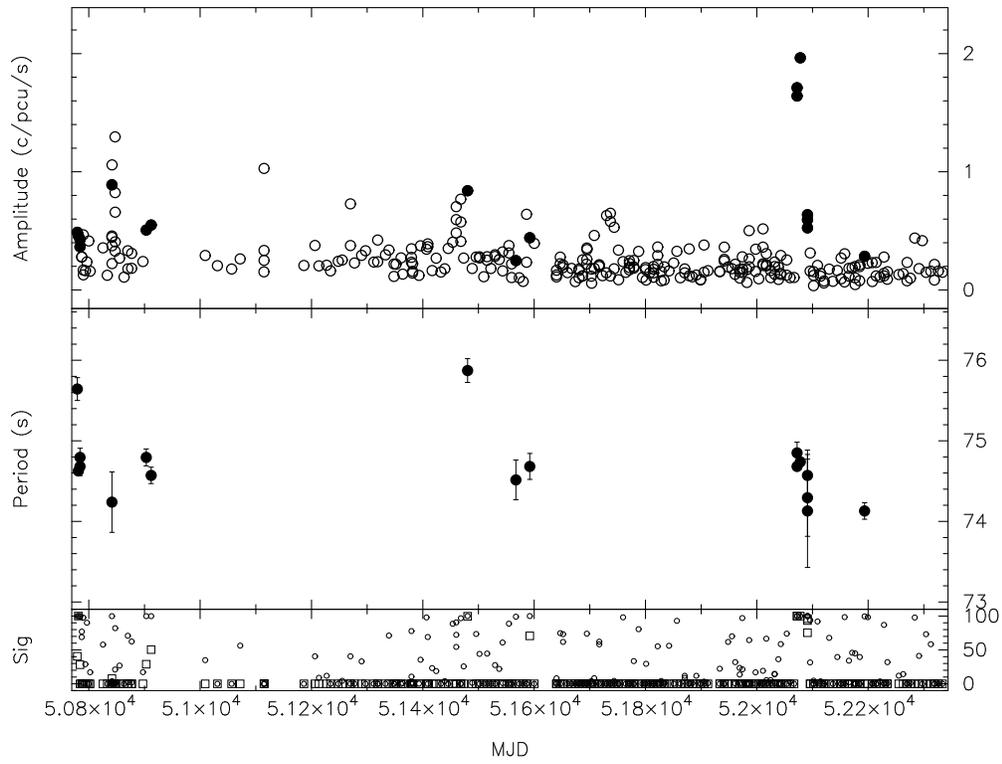}
\caption{AX J0049-729. Pulsed flux and period history.}         
\label{fig:74amp}
\end{figure}\clearpage

\begin{figure}
\includegraphics[width=10cm,angle=-90]{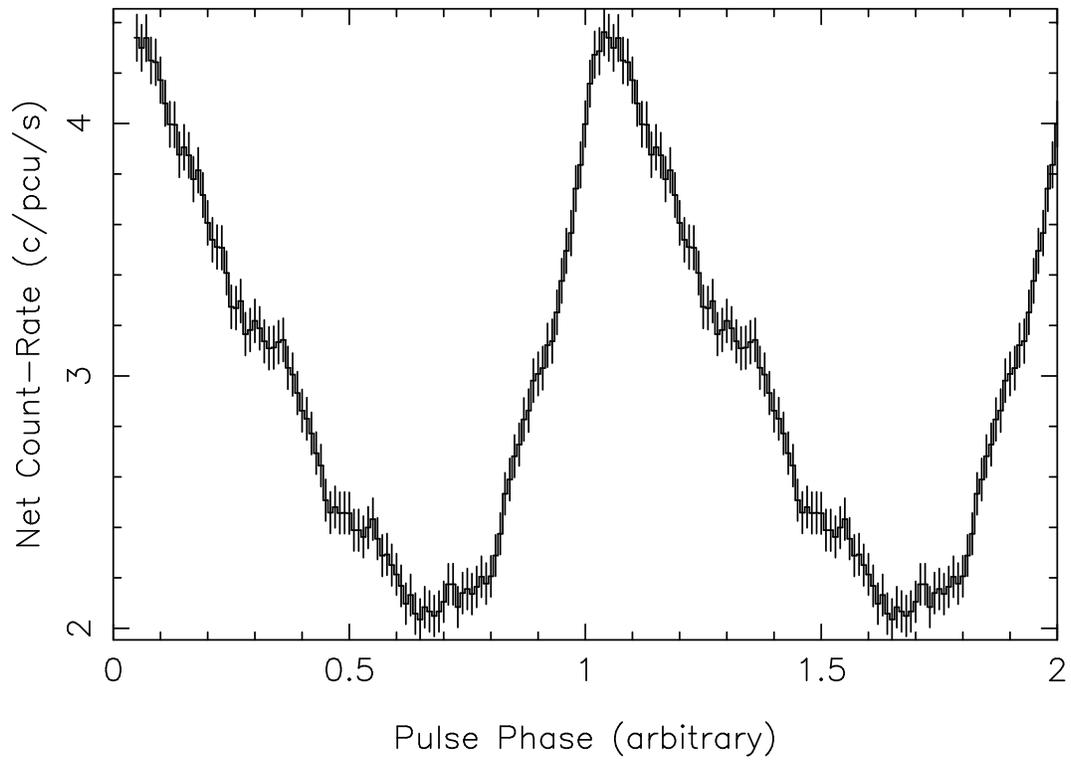}
\caption{Pulse profile of the 74s pulsar AX J0049-729}
\label{fig:74pro} 
\end{figure}\clearpage

\begin{figure}
\includegraphics[width=10cm,angle=-90]{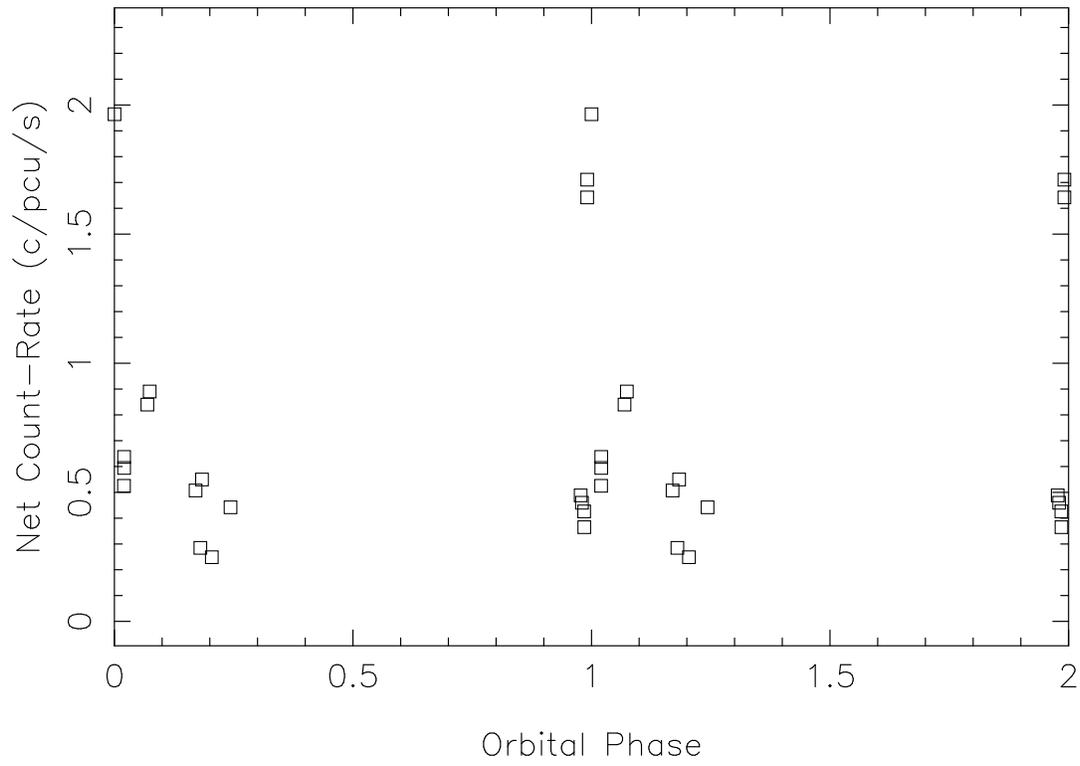}
\caption{AX J0049-729. 99\% significance detections 
of the 74s pulsar folded at 642$\pm$59 days.}
\label{fig:74orbit}
\end{figure}\clearpage
 
\clearpage
 
\begin{figure}
\includegraphics[width=10cm,angle=-90]{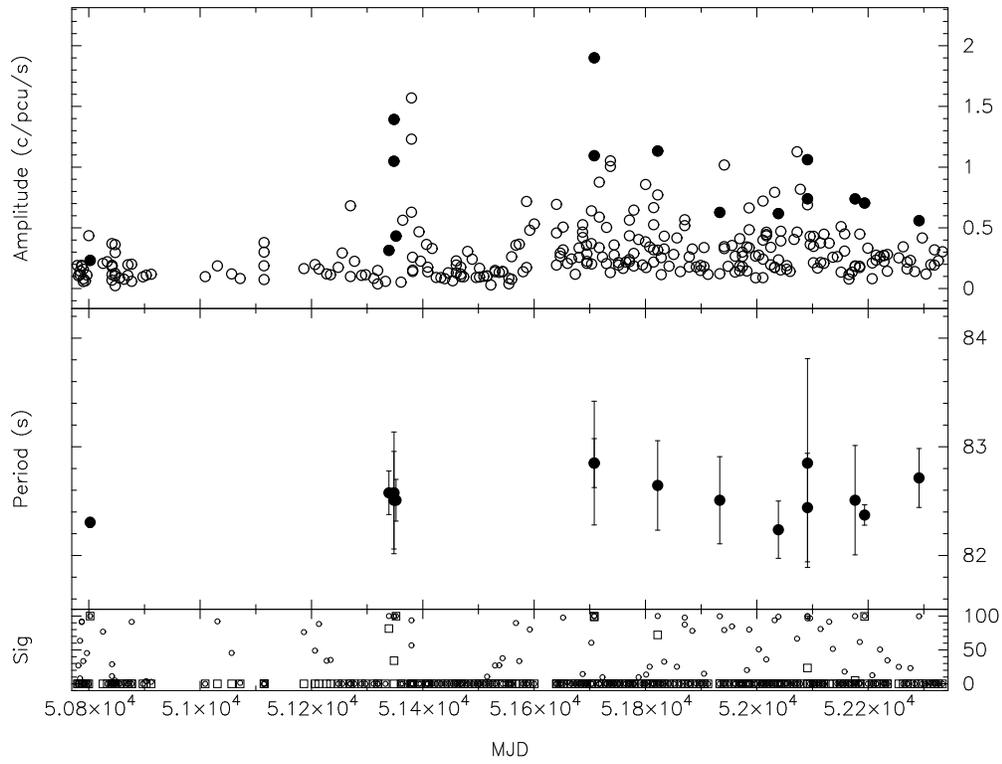}
\caption{XTE J0052-725. Pulsed flux and period history.}  
\label{fig:82amp}
\end{figure}\clearpage

\begin{figure}
\includegraphics[width=10cm,angle=-90]{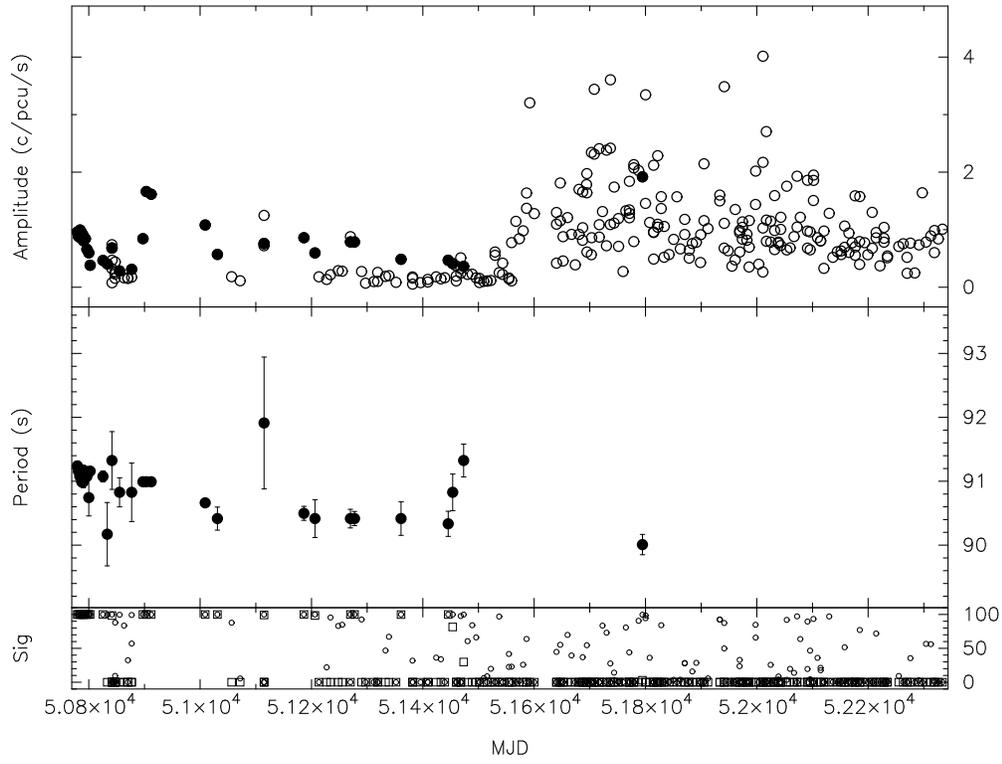}
\caption{XTE 91s (AX J0051-722). Pulsed flux and period history.}
\label{fig:91amp}
\end{figure}\clearpage

\begin{figure}
\includegraphics[width=10cm,angle=-90]{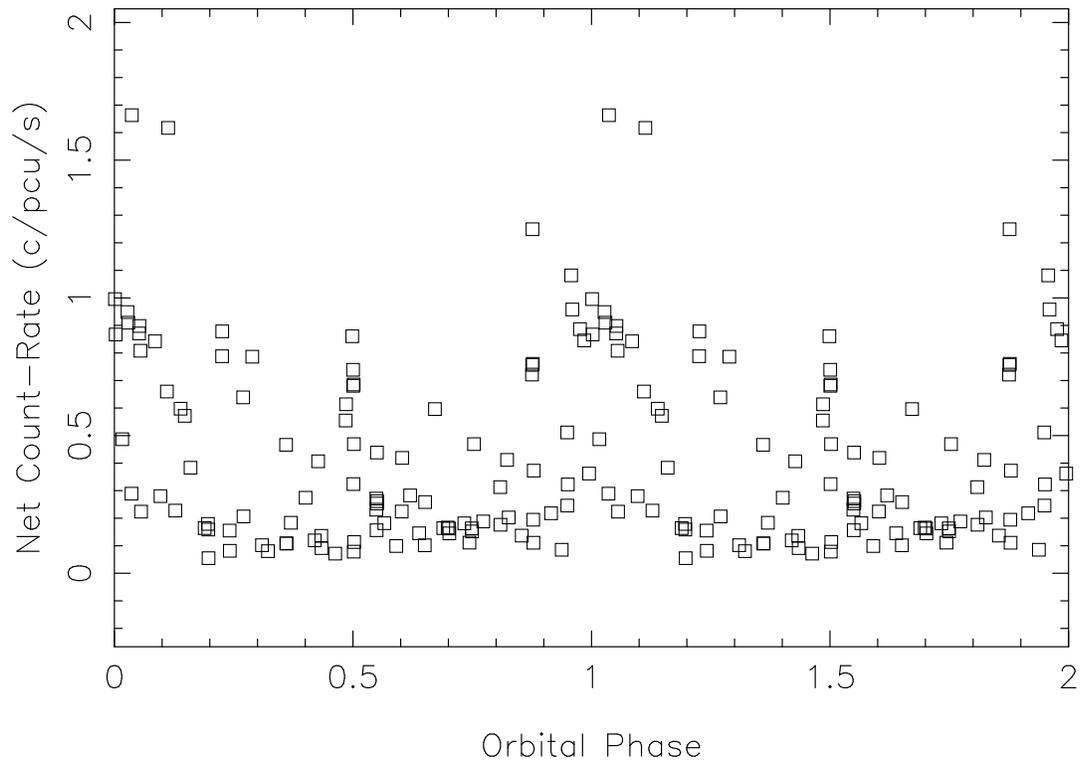}
\caption{Pulsed flux lightcurve for the 91s pulsar AX J0051-722, folded at a period of 115 days}
\label{fig:91fold}
\end{figure}\clearpage

\begin{figure}
\includegraphics[width=10cm,angle=-90]{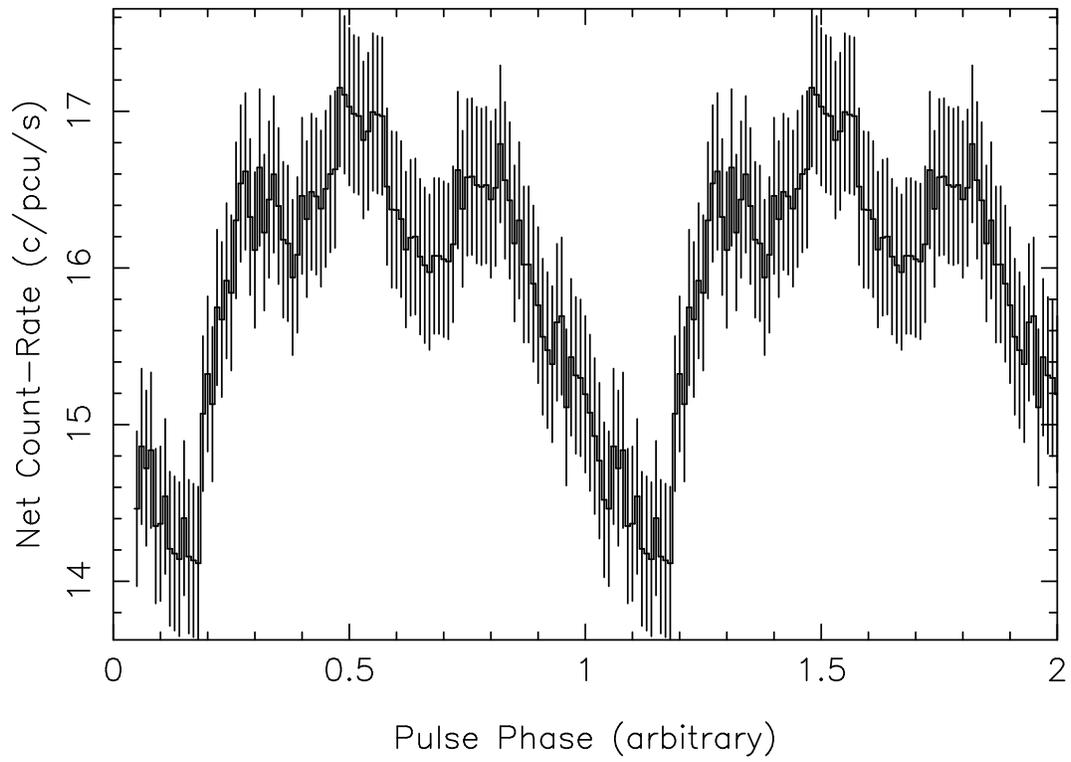}
\caption{Pulse profile for the 91 second pulsar AX J0051-722.}
\label{fig:91pro}
\end{figure}\clearpage

\begin{figure}
\includegraphics[width=10cm,angle=-90]{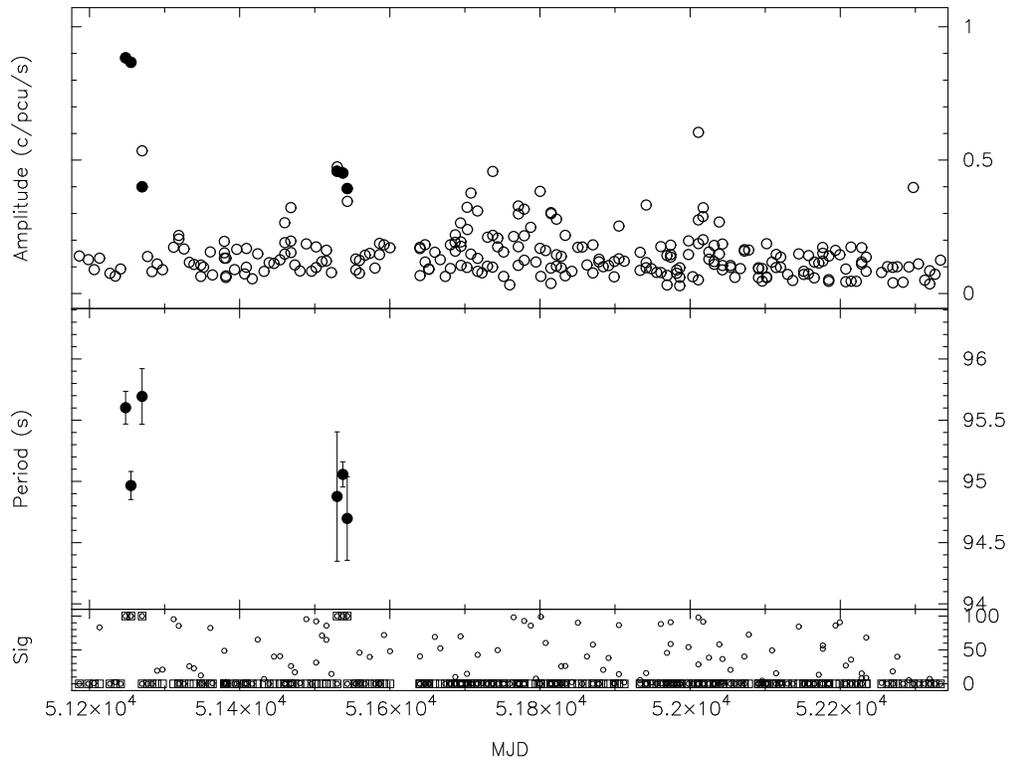}
\caption{XTE SMC95. Pulsed flux and period history.}
\label{fig:95amp}
\end{figure}\clearpage

\begin{figure}
\includegraphics[width=10cm,angle=-90]{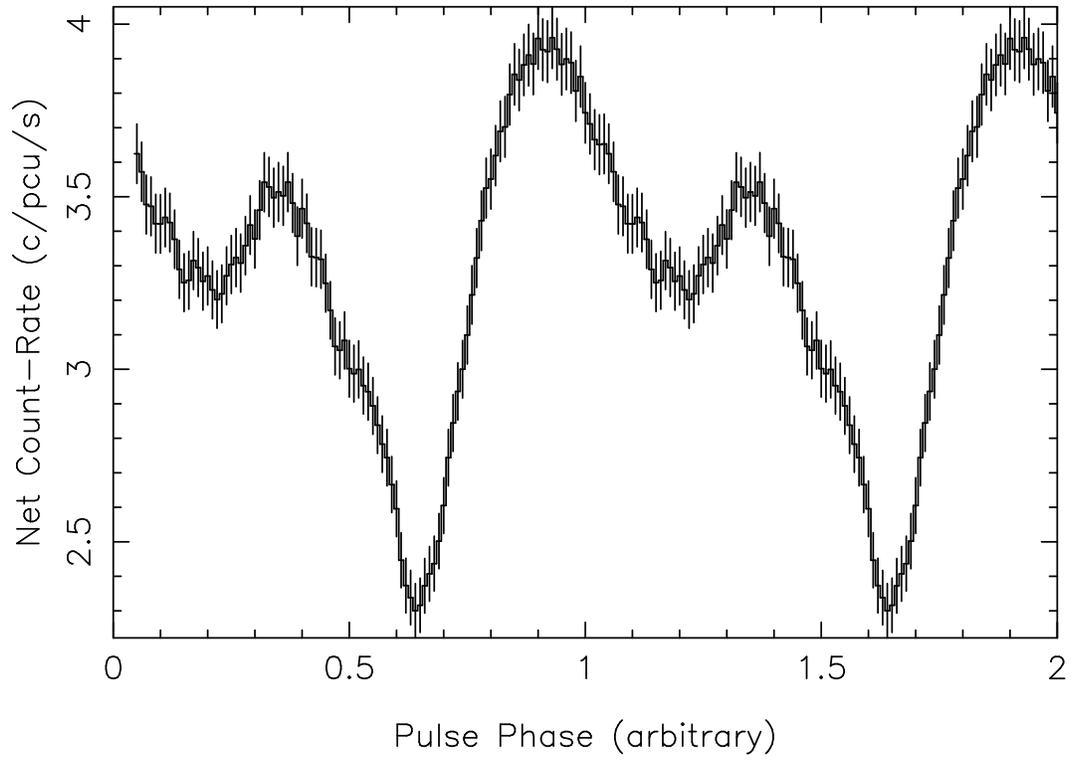}
\caption{Pulse profile for the 95s XTE pulsar.}
\label{fig:95pro}
\end{figure}\clearpage

\begin{figure}
\includegraphics[width=10cm,angle=-90]{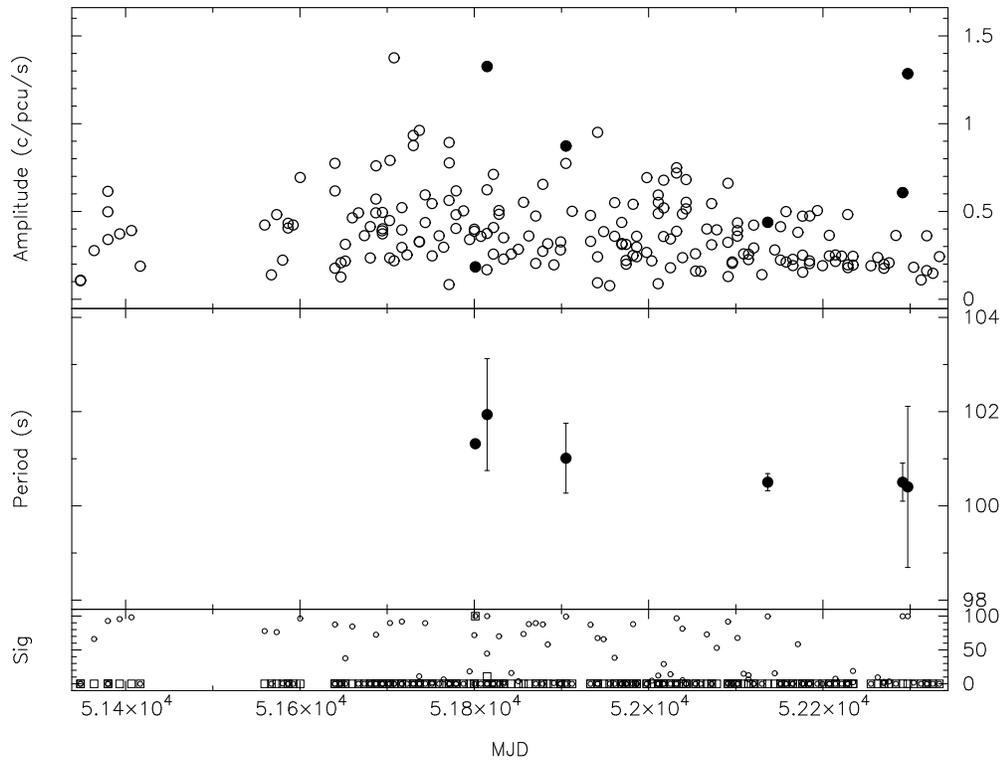}
\caption{AX J0057.4-7325. Pulsed flux and period history.}          
\label{fig:101amp}
\end{figure}\clearpage

\begin{figure}
\includegraphics[width=10cm,angle=-90]{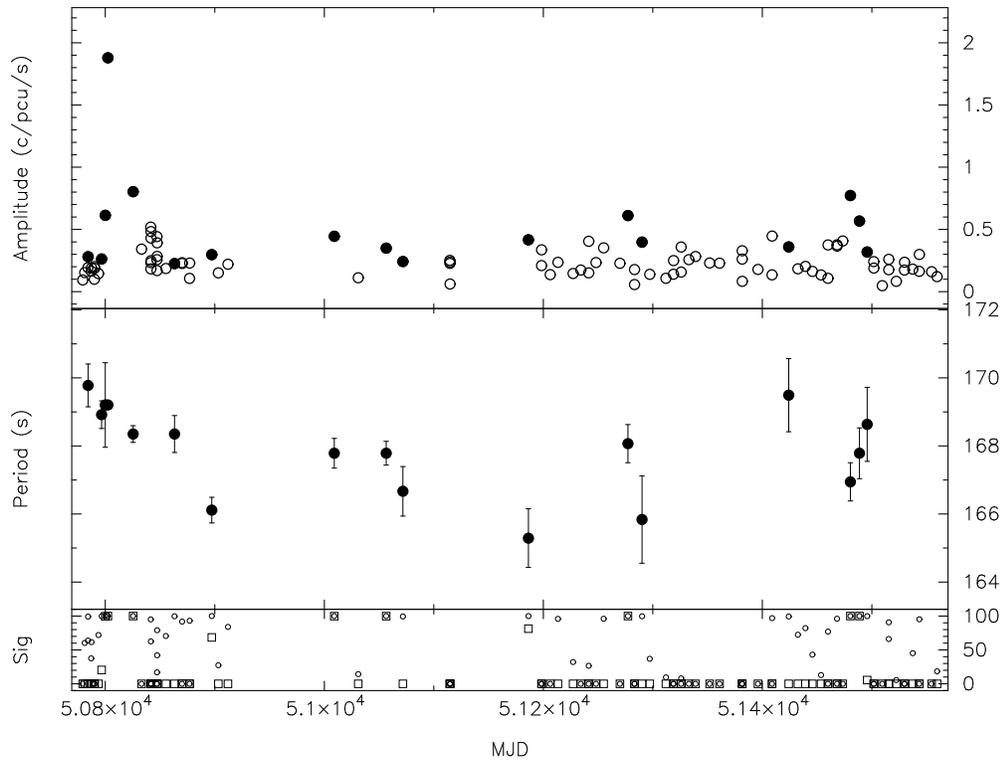}
\caption{XTE J0054-720. Pulsed flux and period history.}
\label{fig:169amp}
\end{figure}\clearpage

\begin{figure}
\includegraphics[width=10cm,angle=-90]{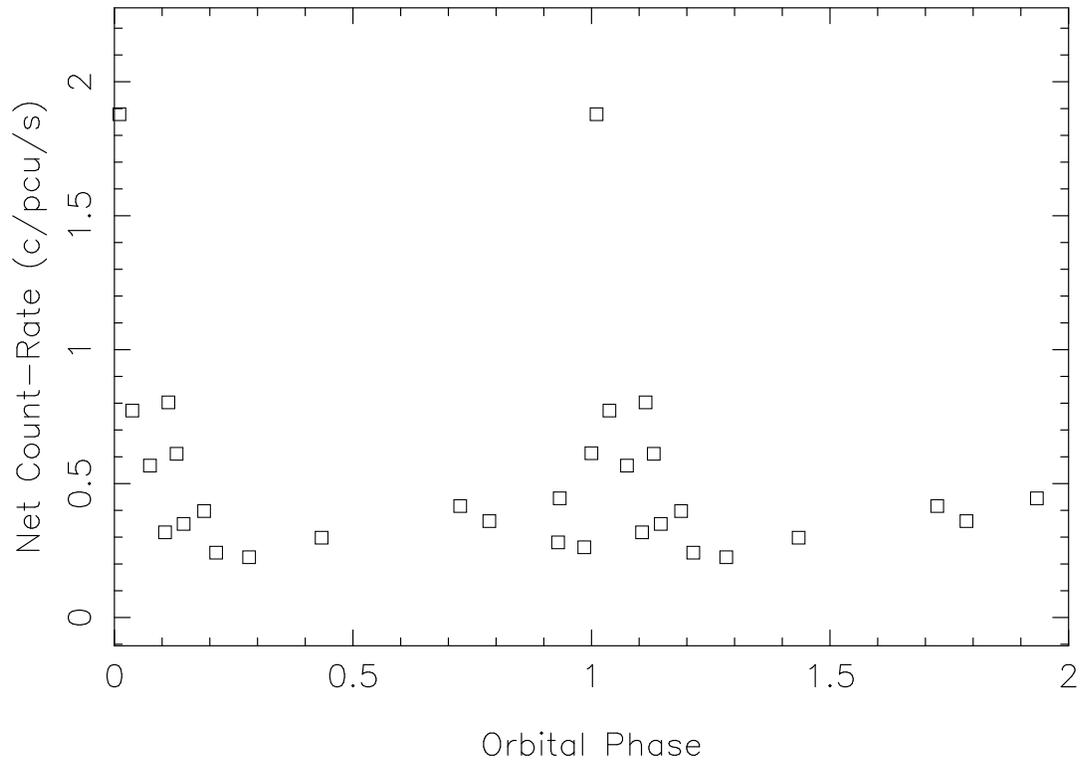}
\caption{Folded 99\% detections of the 169s pulsar XTE J0054-720 at a period of 224 days. }
\label{fig:169fold}
\end{figure}\clearpage

\begin{figure}
\includegraphics[width=10cm,angle=-90]{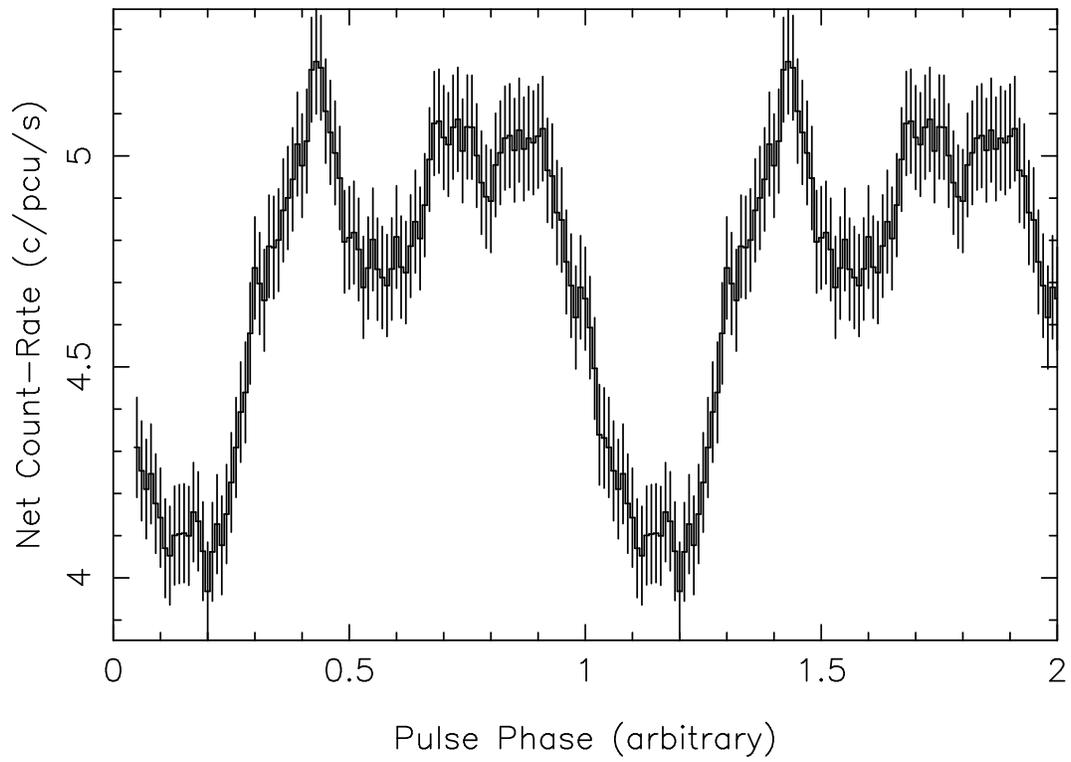}
\caption{Pulse profile for XTE J0054-720 (169s)}
\label{fig:169pro}
\end{figure}\clearpage

\begin{figure}
\includegraphics[width=10cm,angle=-90]{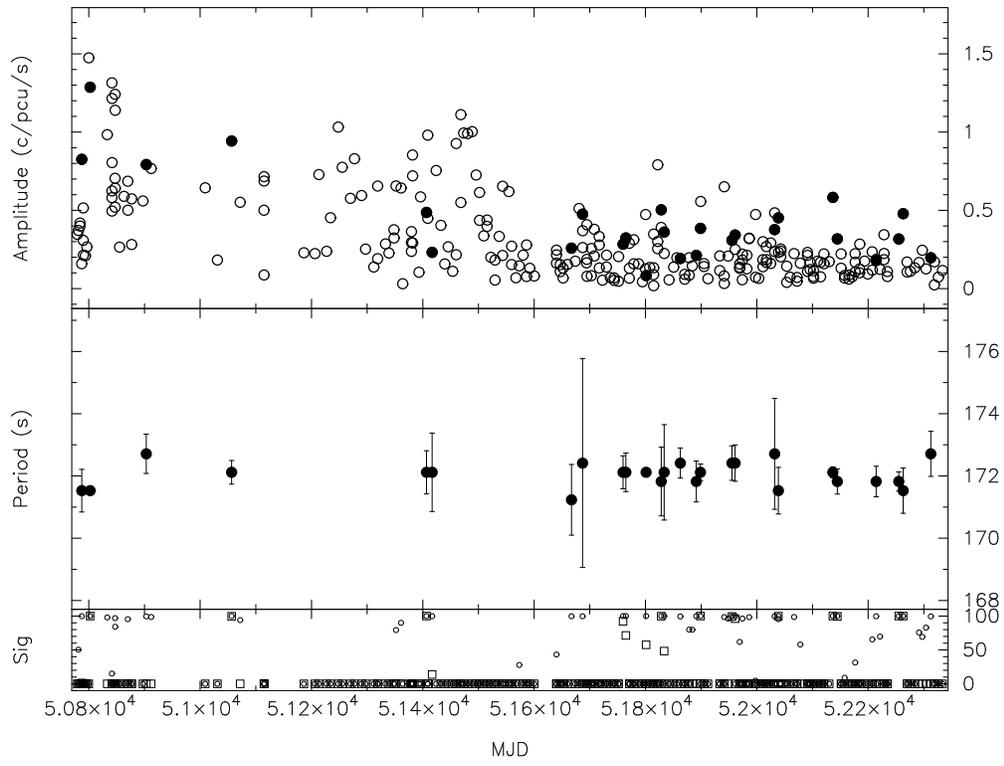}
\caption{AX J0051.6-7311. Pulsed flux and period history.}  
\label{fig:172amp}
\end{figure}\clearpage

\begin{figure}
\includegraphics[width=10cm,angle=-90]{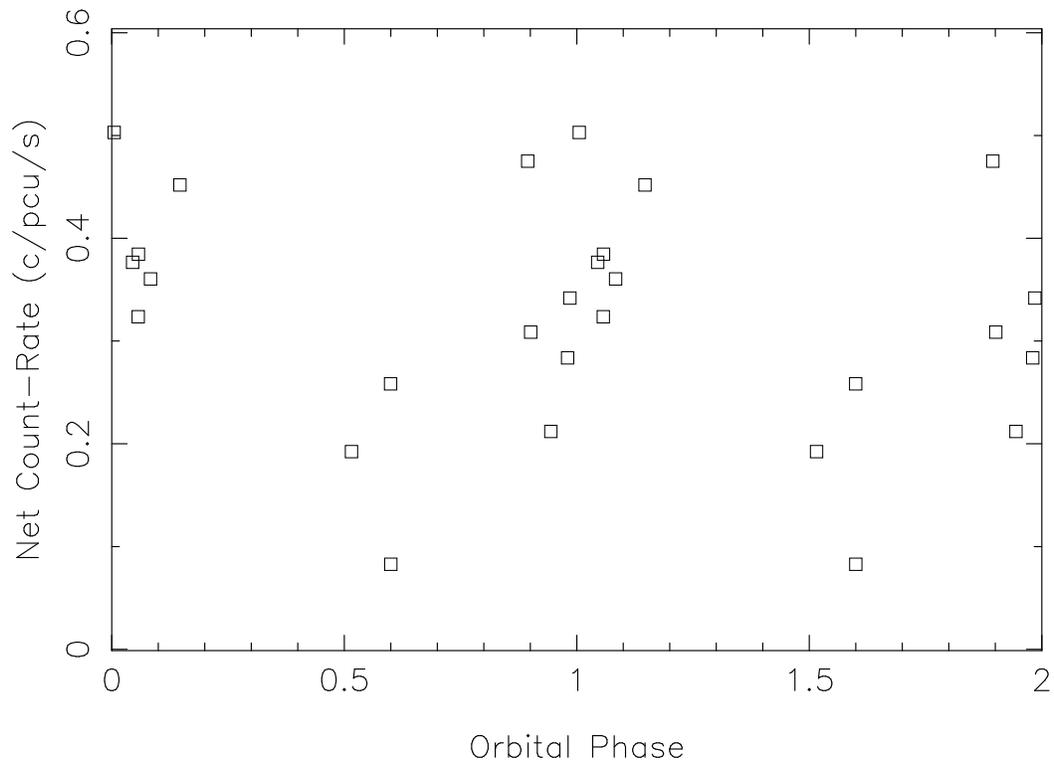}
\caption{Lightcurve for AX J0051.6-7311, 99\% significance detections (MJD 51600-52100) 
folded at a period of 67d.}
\label{fig:172fold}
\end{figure}\clearpage

\begin{figure}
\includegraphics[width=10cm,angle=-90]{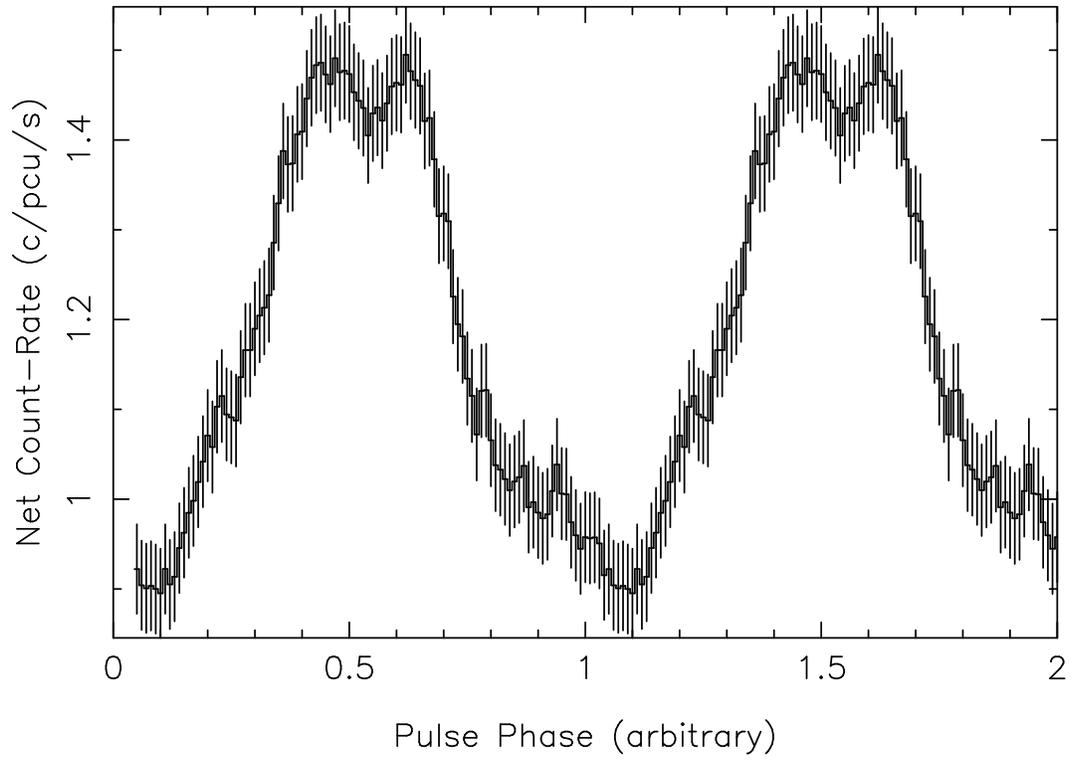}
\caption{Pulse profile for the 172.4 second pulsar AX J0051.6-7311}
\label{fig:172pro}
\end{figure}\clearpage

\begin{figure}
\includegraphics[width=10cm,angle=-90]{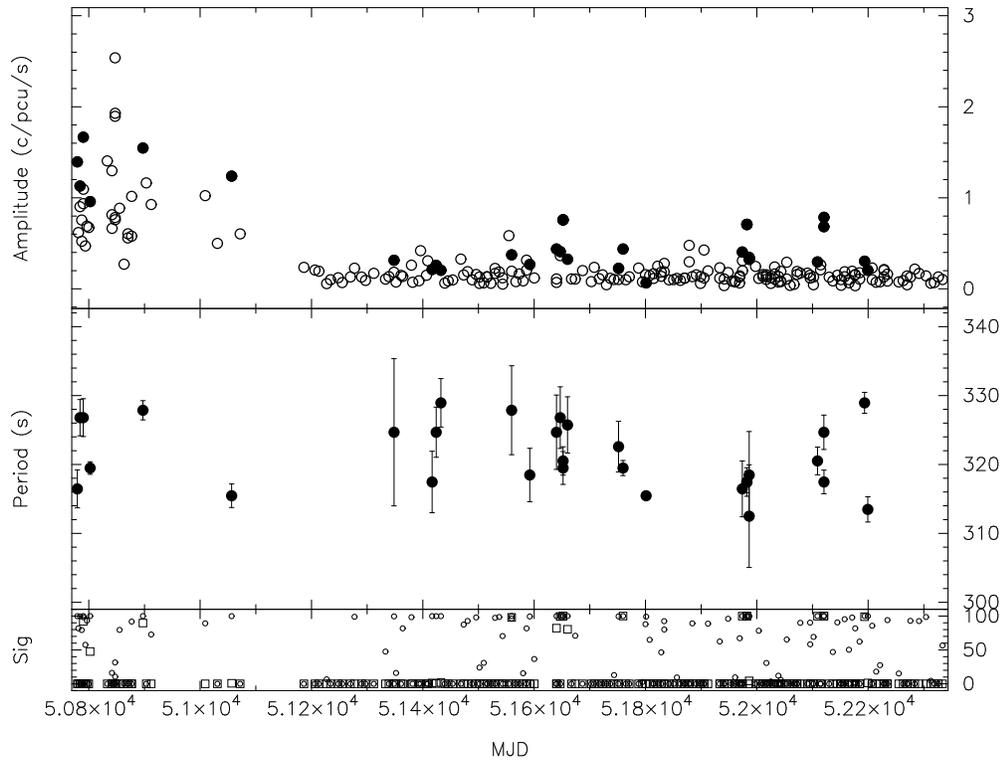}
\caption{RX J0050.8-7316. Pulsed flux and period history.}   
\label{fig:323amp}
\end{figure}\clearpage

\begin{figure}
\includegraphics[width=10cm,angle=-90]{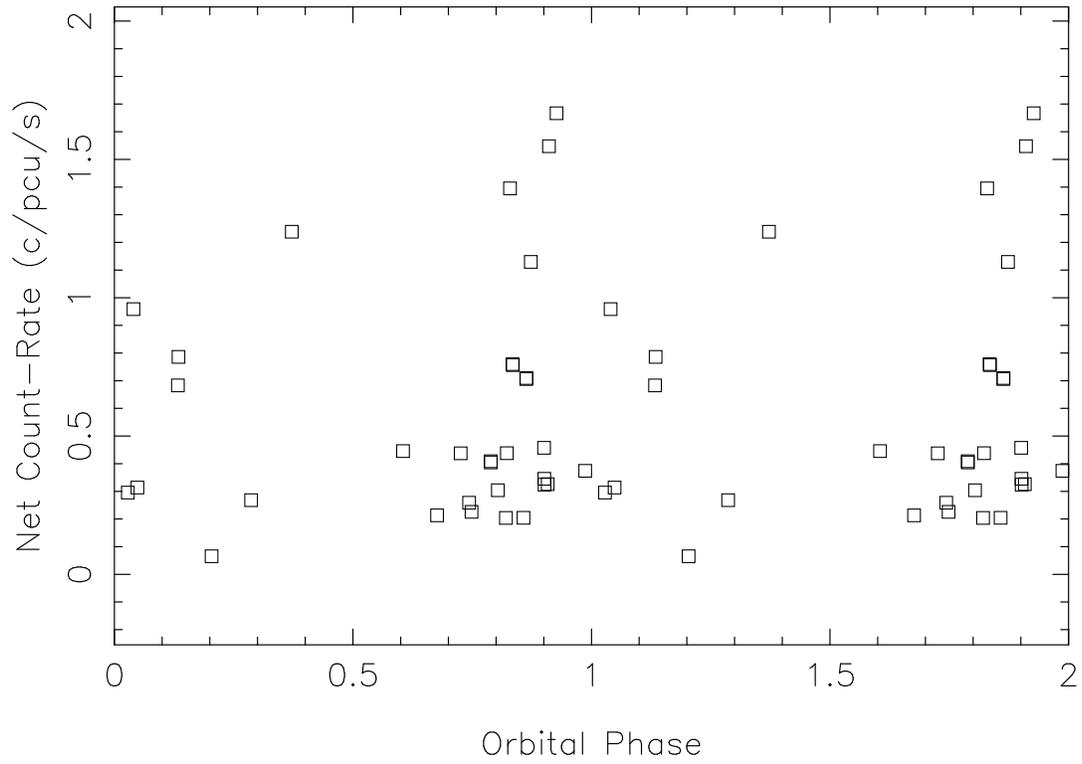}
\caption{Folded lightcurve for RX J0050.8-7316 (323s) at a period of 109 days. The 6 brightest
points are from observations at position 1, the rest are from position 5. (See Table~\ref{tab:positions})}
\label{fig:323fold}
\end{figure}\clearpage

\begin{figure}
\includegraphics[width=10cm,angle=-90]{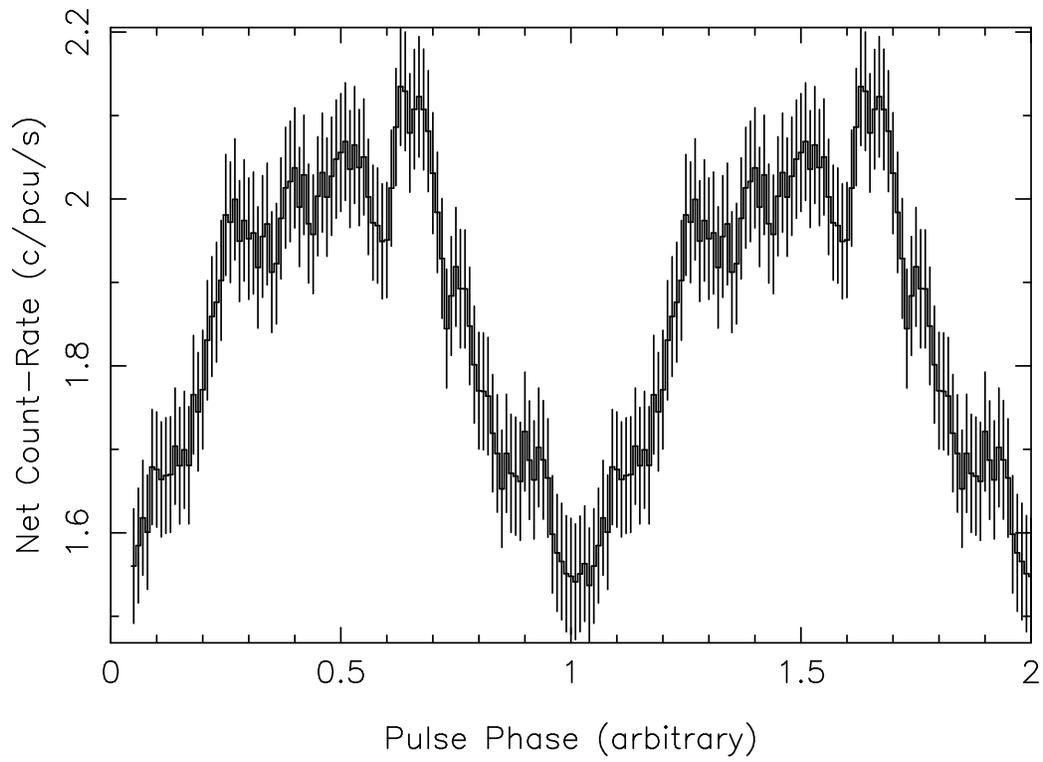}   
\caption{Pulse profile for the 323s pulsar RX J0050.8-7316}
\label{fig:323pro}
\end{figure}\clearpage
 
\clearpage
 
\begin{figure}
\includegraphics[width=10cm,angle=-90]{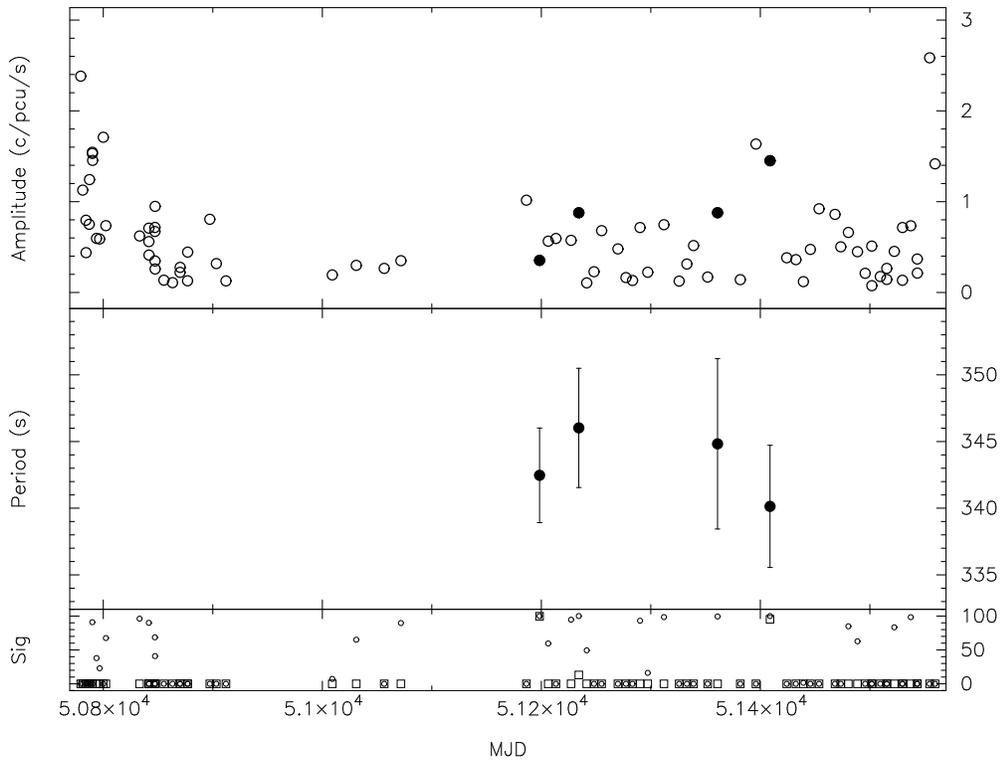}
\caption{AX J0103-722. Pulsed flux and period history.}
\label{fig:348amp}
\end{figure}\clearpage

\begin{figure}
\includegraphics[width=10cm,angle=-90]{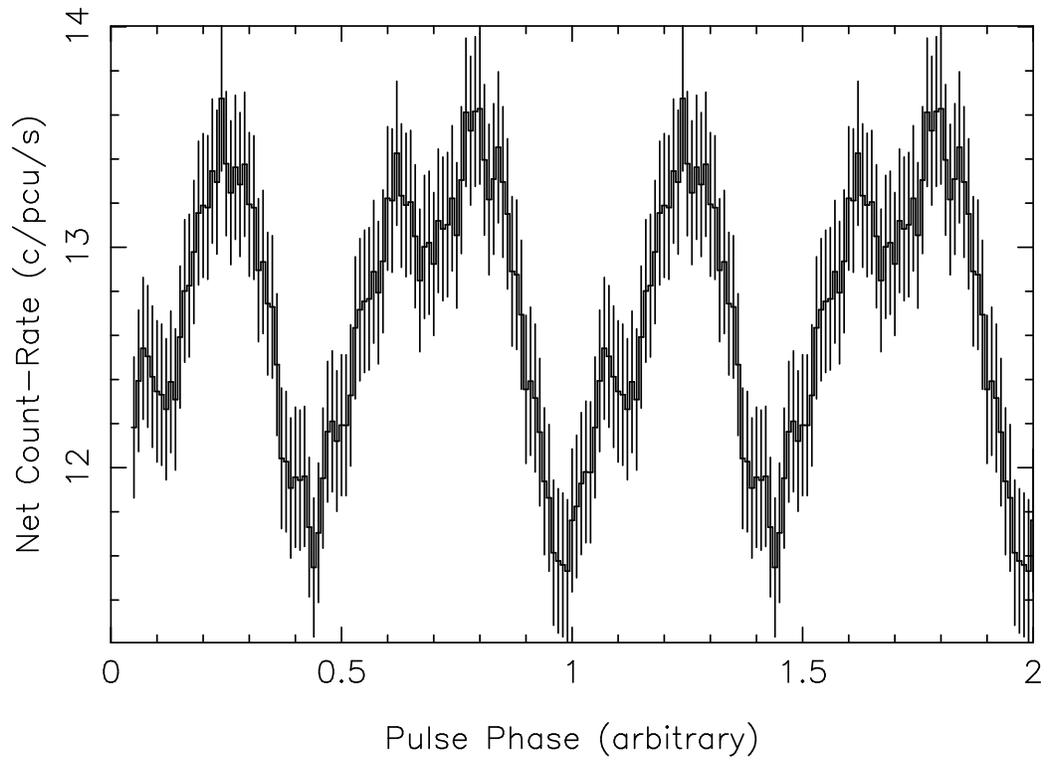}
\caption{Pulse profile for the 348s pulsar AX J0103-722}
\label{fig:348pro} 
\end{figure}\clearpage

\begin{figure}
\includegraphics[width=10cm,angle=-90]{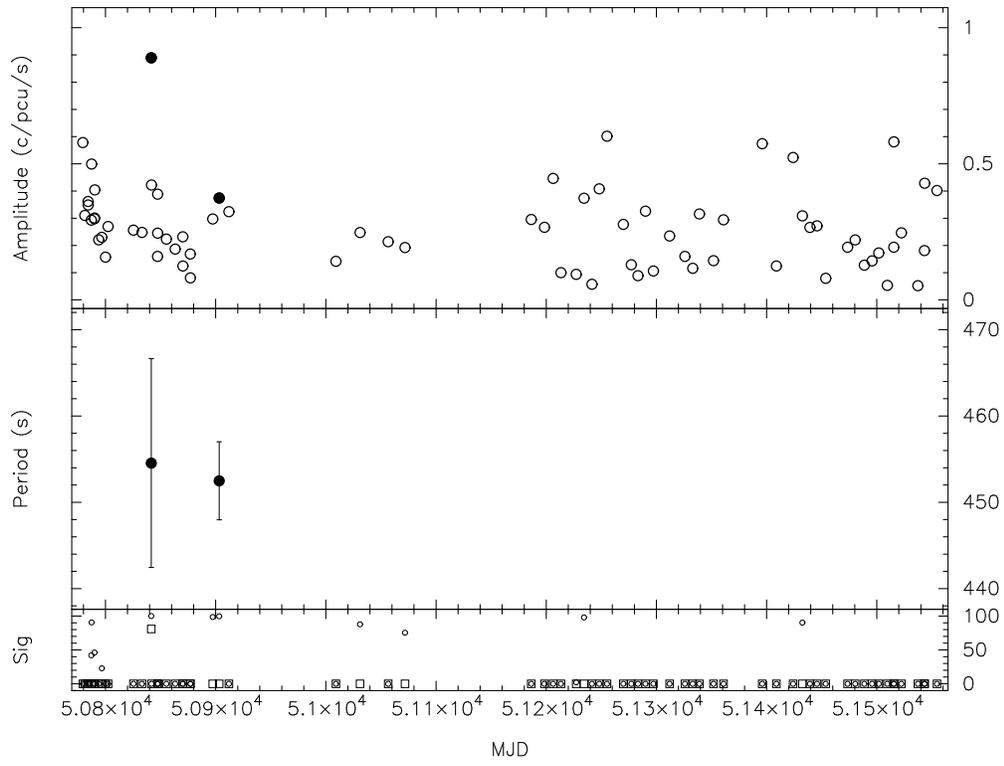}
\caption{RX J0101.5-7211. Pulsed flux and period history.}
\label{fig:455amp}
\end{figure}\clearpage

\begin{figure}
\includegraphics[width=10cm,angle=-90]{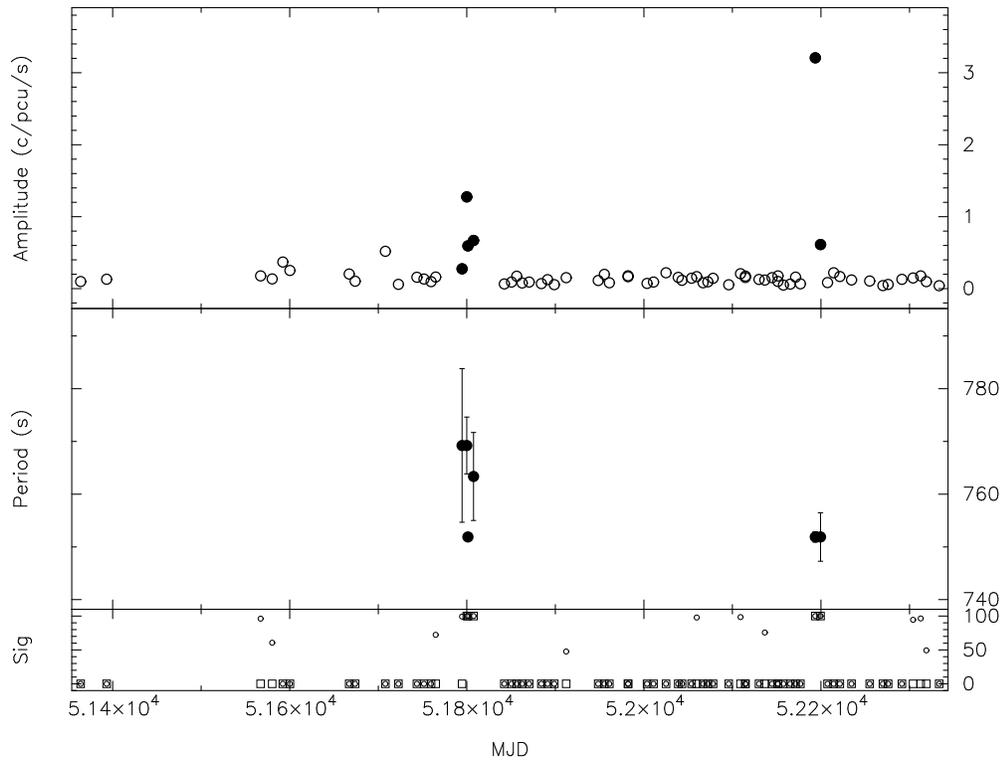}
\caption{AX J0049.4-7323. Pulsed flux and period history. 
Two outbursts exceeding 3$\times 10^{37}$ \ergps\ are separated by 396 days.}
\label{fig:755amp}
\end{figure}\clearpage

\begin{figure}
\includegraphics[width=10cm,angle=-90]{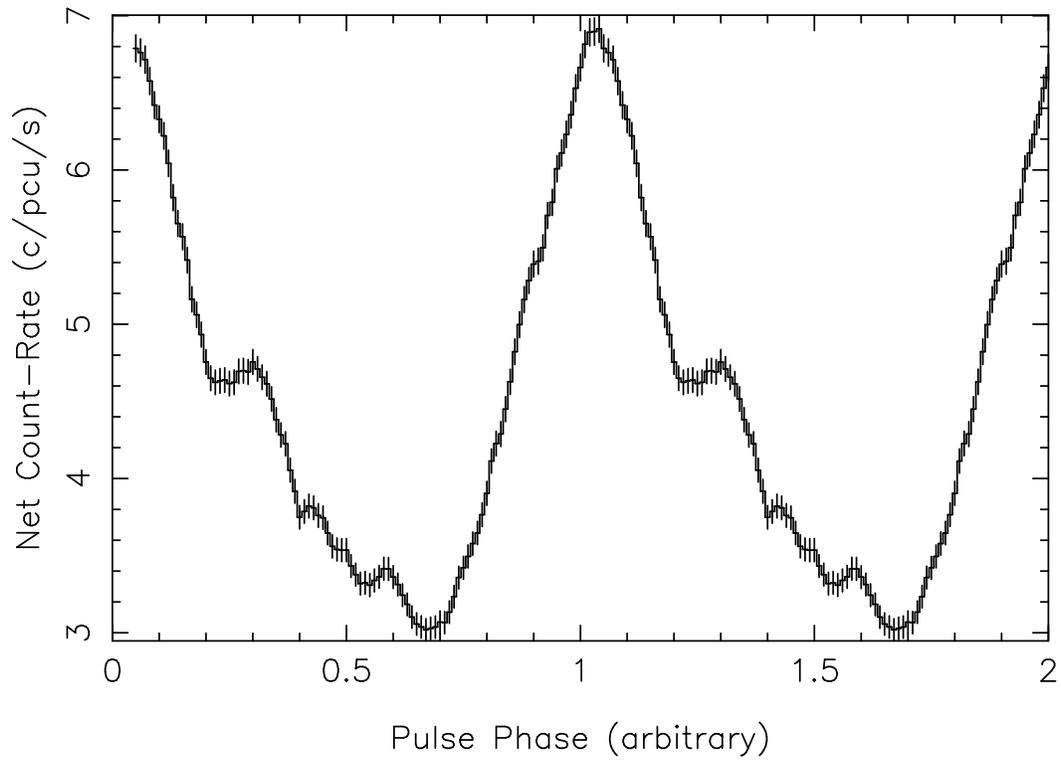}
\caption{Pulse profile for the 751s pulsar AX J0049.4-7323.}
\label{fig:755pro}
\end{figure}\clearpage

\begin{figure}
\includegraphics[width=10cm,angle=90]{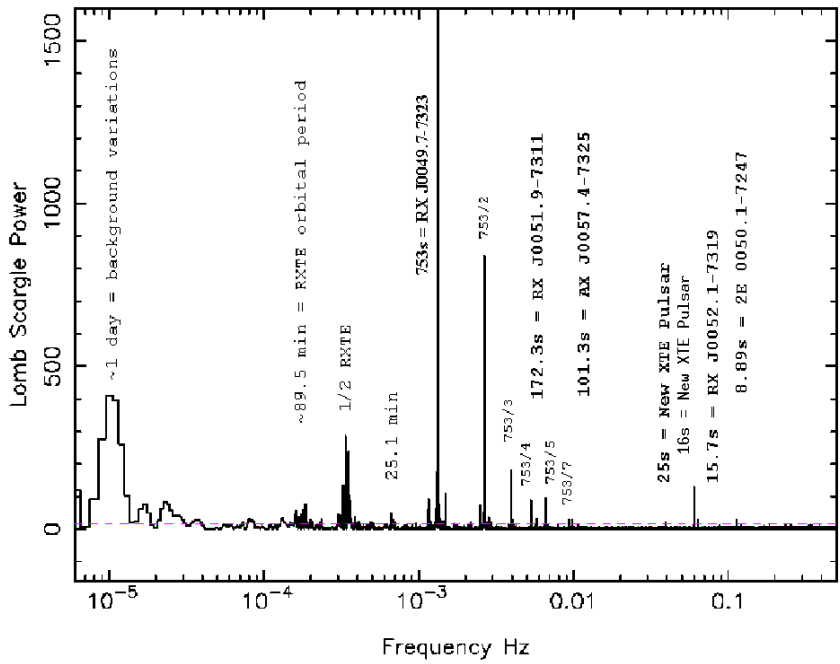}
\caption{Periodogram for the deep (2 day) observation shows 7 pulsars, 
 two of which were new discoveries. From left: RX J0049.7-7323, RX J0051.9-7311, AX J0057.4-7325, XTE 51sec, 
XTE16.6sec, RX J0052.1-7319, 2E 0050.1-7247}
\label{fig:longls}
\end{figure}\clearpage

\begin{figure}
\begin{center}
\includegraphics[width=10cm,angle=-90]{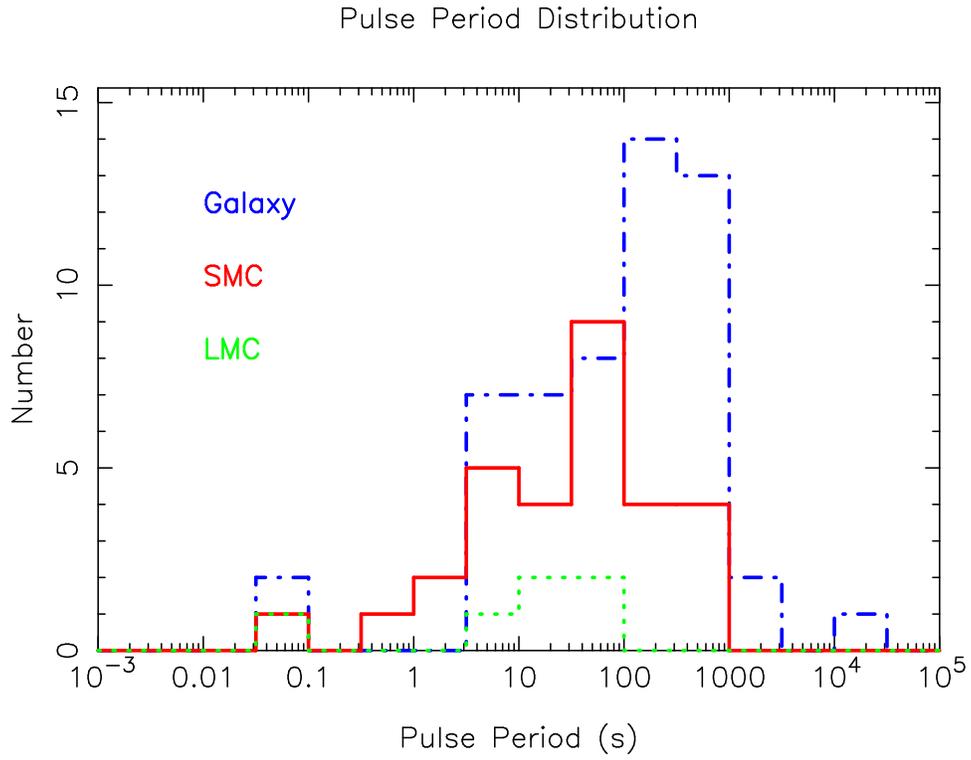}
\caption{Distribution of pulse periods in the SMC, LMC and the Galaxy. The SMC population 
appears to be shifted to shorter periods, the significance of this effect is 97\% 
according to the K-S test. }
\label{fig:dist_smclmcgal}
\end{center}  
\end{figure}\clearpage

\begin{figure}
\includegraphics[width=10cm,angle=-90]{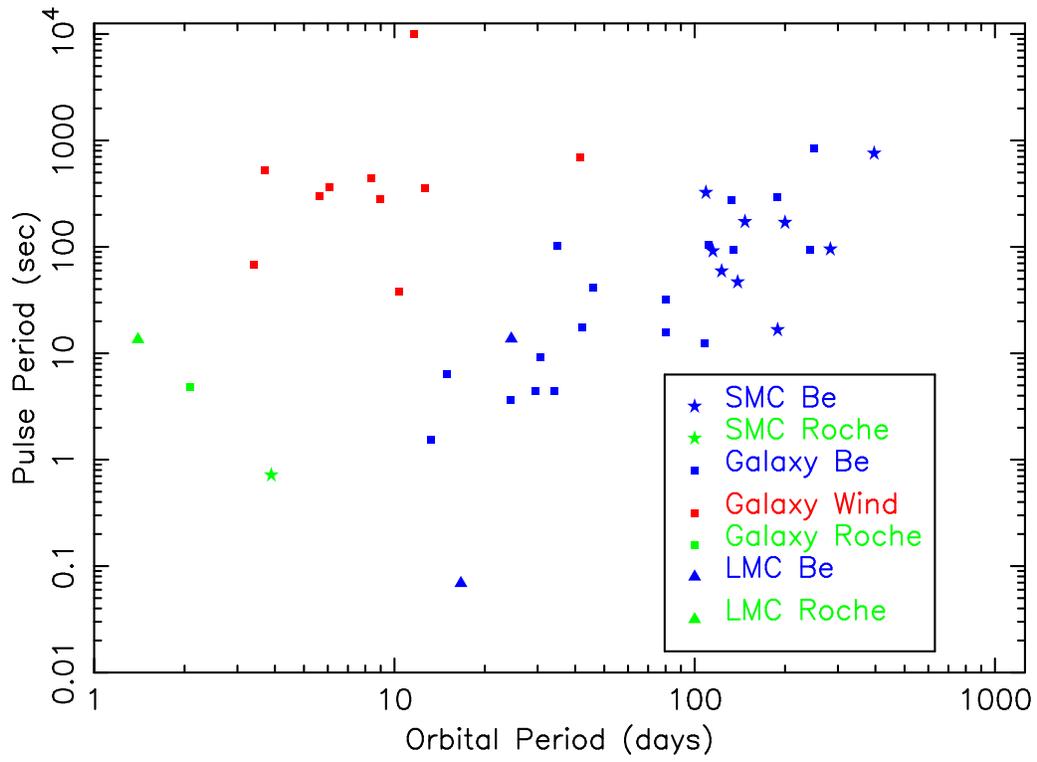}
\caption{The $P_{pulse}/P_{orbit}$ diagram for HMXBs in the Galaxy and Magellanic Clouds,
the SMC orbital periods are the candidate periods presented in this paper.}
\label{fig:newcorbet}
\end{figure}\clearpage

\begin{figure}
\includegraphics[width=10cm,angle=0]{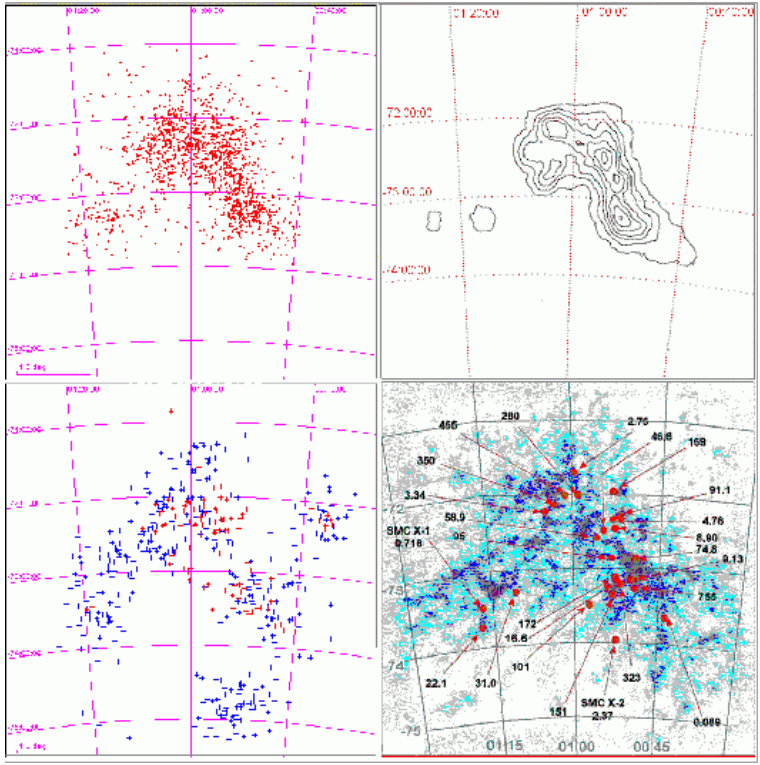}
\caption{Spatial distributions (clockwise from top left) 1. Emission line stars, 2. Isodensity 
contours for stars aged 8-12.2 My, 3. Neutral hydrogen map, with positions of X-ray pulsars 
superimposed, 4. \rosat X-ray Sources, blue +'s= PSPC, red +'s= HRI point sources. Credits: 1. 
ALADIN/\citet{MA93}, 2. adapted from \citet{Maragoudaki01}, 3. HI map adapted from 
\citet{S99}, 4. ALADIN/\citet{Haberl00,Sasaki00}}.
\label{fig:spatial}
\end{figure}\clearpage



\begin{table}
\caption{Pointing Positions for SMC Monitoring.}
\begin{tabular}{lrcll}
\tableline
 Position   &  $N_{obs}(N_{sec})$ &  Dates (MJD)  &  $RA(\degr)$  &  $Dec(\degr)$   \\
\tableline 
 1a         &    9 (13)           & 50779 - 50802 &  13.033      &  -72.429    \\
 1b         &   1 (1)             &     50825     &  12.767     &  -72.229    \\
 1c         &   18 (31)           & 50833 - 51115 &  13.728      &  -72.445     \\
 1          &   41 (54)           & 51186 - 51555 &  13.471       &  -72.445        \\
 2          &   4 (6)                 & 51198 - 51326 &  16.25        &  -72.1          \\
 3          &   3 (4)                & 51220 - 51310 &  18.75        &  -73.1          \\
 4          &   6 (4)                & 51348 - 51417 &  12.686       &  -73.268        \\
 5          &   45 (70)               & 51560 - 52333 &  12.5         &  -73.1          \\ 
\end{tabular}
\medskip

\tablecomments{Column $N_{obs}(N_{sec})$ indicates 
the number of whole observations followed by the 
number of separate sections into which these were split for analysis. Coordinates are
epoch 2000.}
\label{tab:positions} 
\normalsize
\end{table}


\begin{table}
\caption{\rxte PCA Pointed Observations of SMC X-2.\label{tab:x2flux}}
\begin{tabular}{lllll}
\hline
 Obs.   &  Date      & Period & Flux           &  Pulsed Flux  \\
        & (2000)     & (sec)  & \fluxpcu       & \fluxpcu     \\
\hline
 1$^{\dag}$&  Jan 24 & 2.3728(1)  & 0.03          & 0.16   \\
 2      &  Jan 30    & 2.37210(5) & 10.04          & 2.91       \\
 3      &  Feb 6     & 2.3717(1)  & 11.32          & 1.34       \\
 4      &  Feb 12    & 2.37281(1) & 16.15          & 1.44       \\
 5      &  Feb 18    & 2.371815(5)  & 26.49          & 11.97    \\
 6      &  Feb 26    & 2.371787(5)  & 25.23          & 7.99     \\
 7      &  Apr 6     & 2.372217(5)  & 8.45           & 1.96     \\
 8a     &  Apr 12-13 & 2.37157(1)  & 7.14           & 1.86      \\
 8b$^*$ &  Apr 12    &             & 5.2            & -         \\
 9      &  Apr 18    & 2.3711(1)   & 7.02          & 1.62       \\
 10$^*$ &  Apr 22-23 & 2.371859(5)  & 3.23         & 1.19       \\
\end{tabular}
\medskip 
\tablecomments{$^*$ Observations centered on SMC X-2, previously appeared in \citet{Corbet01_x2} 
Table 1 as numbers 2 \& 3. All other observations
 were centered on RA = $0^h50^m$, Dec = $-73\degr$6\arcmin.
 \dag Observation 1 has unreliable flux calibration due to a telemetry error.}

\normalsize
\end{table}
 

\begin{table}   
 \caption{Spectral fits for SMC X-2}
 \label{tab:x2spec}
 \begin{tabular}{lllccc}
\tableline
Obs                   &   5                          & 8b                     & 10            \\
\tableline
$N_{H}\times10^{22}$  & 1.6 $\pm$ 0.3                & 1.6$\pm$0.8            & 3.3$\pm$0.6   \\
$\alpha$              & 1.17$\pm$0.02                & 0.70$\pm$0.04          & 1.0$\pm$0.03  \\
$E_{c}$               & 12.4$\pm$1                   &                        &        \\
$E_{f}$               & 50$\pm$12                    &                        &         \\
$\sigma Fe$           & 0.8$\pm$0.2                  &                        &        \\
norm $Fe$             & (2.9$\pm$1)$\times 10^{-10}$ &                        &         \\
$flux_{2-10}$ & 1.5$\times 10^{-10}$         & 6.93$\times 10^{-11 }$ & 5.71$\times 10^{-11 }$ \\
$flux_{10-20}$& 1.6$\times 10^{-10}$         &                        &  \\
$L_{X}$   & 4.4$\times 10^{38}$          &                 &          \\
$\chi_{\nu}{^2}$      & 1.25                         & 1.52                   &  1.12    \\
\end{tabular}
\medskip 
\tablecomments{Units are keV for $E_{c}$, $E_{f}$, $\sigma Fe$.
Flux units \fluxerg. Luminosity units (unabsorbed) erg s$^{-1}$ at 65 kpc, 
collimator response 0.33. 
Parameters for observations 8b and 10 are taken from \citet{Corbet01_x2}.}
\normalsize
\end{table}


\begin{table}
\caption{Pulsars not detected with \rxtec. 
\label{tab:nondetect}}
\begin{tabular}{llll}
\tableline
Pulsar      & Period (s) & \fluxpcu & Ref. \\     
\tableline
RX J0059.2-7138 & 2.8   & $<$0.40 &  \cite{Hughes1994} \\
AX J0105-722    & 3.34  & $<$0.16 &  \cite{IAU7028} \\
AX J0049-732    & 9.13  & $<$0.10 &  \cite{IAU7040} \\
RX J0117.6-7330 & 22    & $<$0.29 & \cite{IAU6305} \\ 
AX J0058-7203   & 280   & $<$0.17 & \cite{Yokogawa98a} \\
\end{tabular}
\tablecomments{Upper limits on the \rxte PCA pulsed-flux for 5 previously known
Be/X-ray pulsars, which were not detected during our survey.}  
\end{table}

\input{tab5}

\input{tab6}

\input{tab7}

\end{document}

%% file: tab5.tex
\begin{deluxetable}{llllllllllllllll}
\rotate
\tablewidth{0pt}
\tabletypesize{\scriptsize}
\tablecaption{Measured X-ray properties for SMC pulsars.\label{tab:catalogue}}
\scriptsize

\tablehead{
\colhead{P$_{pulse}$}& 
\colhead{P$_{orbit}$ } & 
\colhead{T${_0}$  } &
\colhead{Fp$_{min}$ }& 
\colhead{Fp$_{max}$ }& 
\colhead{f$_{X}^{2 - 10}$ }& 
\colhead{f$_{X}^{10 - 20}$ } & 
\colhead{$\alpha $ } & 
\colhead{n$_{H }$ } & 
\colhead{E$_{cut}$ }& 
\colhead{E$_{fold}$ }& 
\colhead{$\sigma $Fe } & 
\colhead{EW(Fe)} & 
\colhead{$\chi^{2}$  } & 
\colhead{R(pos)} & 
\colhead{L$_{x}^{2 - 10}$}\\

\colhead{ s }& 
\colhead{ days} & 
\colhead{ MJD } &
\colhead{ }& 
\colhead{ }& 
\colhead{ }& 
\colhead{ } & 
\colhead{ } & 
\colhead{\quad} & 
\colhead{keV}& 
\colhead{keV}& 
\colhead{eV} & 
\colhead{eV} & 
\colhead{ } & 
\colhead{ } & 
\colhead{  }\\

}
\startdata
{2.374}& 
& 
&
1.29& 
12.41& 
14.95& 
16.0& 
1.17& 
1.63& 
12.4& 
50& 
0.75& 
148& 
1.25& 
0.33 (5)& 
21.5 \\
{4.78}& 
&
& 
1.63& 
2.29& 
13.47& 
9.08& 
1.7& 
2.5& 
15& 
12& 
0.78& 
196& 
0.86& 
1 (*)& 
7.241 \\
{16.6}& 
$^1$189$\pm $18& 
51393&
0.27& 
0.97& 
& 
& 
& 
& 
& 
& 
& 
& 
& 
& 
 \\
{31.0}& 
&
& 
-& 
0.90& 
8.905& 
10.99& 
0.968& 
1.46& 
12.23& 
29& 
0.42& 
122 \quad & 
0.99& 
0.67 (3)& 
6.845 \\
{46.6}& 
$^2$139$\pm $6& 
50779&
0.19& 
0.92& 
& 
& 
& 
& 
& 
& 
& 
& 
& 
& 
 \\
{51}& 
&
& 
& 
& 
& 
& 
& 
& 
& 
& 
& 
& 
& 
& 
 \\
{58.97}& 
$^3$123$\pm $1& 
50841&
0.18& 
2.76& 
7.3677& 
5.84& 
1.26& 
2.9& 
7.6& 
18& 
no& 
-& 
0.6& 
0.98 (1c)& 
4.257 \\
{74.7}& 
$^4$642$\pm $59& 
52078&
0.24& 
1.96& 
2.157& 
1.90& 
1.219& 
1.395& 
16.24& 
2.5& 
0.54& 
182 \quad & 
0.61& 
0.75 (5)& 
1.491 \\
{82.4}& 
&
& 
0.24& 
1.90& 
& 
& 
& 
& 
& 
& 
& 
& 
& 
& 
 \\
{91.1}& 
$^5$115$\pm $5& 
50784&
0.28& 
1.92& 
& 
& 
& 
& 
& 
& 
& 
& 
& 
& 
 \\
{95s}& 
$^6$283$\pm $8& 
51248&
0.39& 
0.89& 
2.375& 
2.03& 
1.26& 
0.72& 
15& 
12& 
weak& 
-& 
0.6& 
1 (1*)& 
1.2 \\
{101.4}& 
&
& 
0.18& 
1.33& 
& 
& 
& 
& 
& 
& 
& 
& 
& 
& 
 \\
{169}& 
$^7$200$\pm $40& 
50800&
0.22& 
1.88& 
& 
& 
& 
& 
& 
& 
& 
& 
& 
& 
 \\
{172.4}& 
$^8$147$\pm $24& 
51694&
0.08& 
1.29& 
1.005& 
0.55& 
1.251& 
0.04& 
12& 
5& 
0.51& 
536& 
0.83& 
0.86 (5)& 
0.5563 
\\
{323}& 
$^9$109$\pm $18& 
51651&
0.06& 
1.67& 
1.95& 
1.46& 
1.45& 
4.6& 
13& 
7& 
0.56& 
264& 
0.69& 
0.83 (5)& 
1.1137 \\
{343}& 
&
& 
0.35& 
1.45& 
& 
& 
& 
& 
& 
& 
& 
& 
& 
& 
 \\
{455}& 
&
& 
0.37& 
0.88& 
& 
& 
& 
& 
& 
& 
& 
& 
& 
& 
 \\
{755}& 
$^{10}$396$\pm $5& 
51800&
0.27& 
3.20& 
3.14& 
2.04& 
1.55& 
1.8& 
12.16& 
14& 
0.6& 
330 \quad & 
0.85& 
0.71 (5)& 
2.387 \\
\enddata
\medskip
\tablecomments{
Explanation of columns is given in the text.\\
Notes on period determination: \\
(1) Period determined from 5 evenly spaced faint outbursts. \\
(2) From 6 outbursts.
(3) From 4 very bright and well observed outbursts.
(4) From 3 outbursts, 1 bright, 2 faint.
(5) 2 bright outbursts and up to 6 faint ones.
(6) From 2 bright outbursts.
(7) Up to 4 faint outbursts, uncertain.
(8) Numerous detections. Period determined from a string of 6 evenly spaced brighter outbursts.
(9) Many detections with weak pattern.
(10) 2 very bright outbursts.
} 
\normalsize
\end{deluxetable}

%% file: tab6.tex
\begin{table*}[htbp]
\scriptsize
\caption{List of X-ray Pulsars in the SMC.}
\label{tab:pulsars}
\begin{tabular}
{|p{50pt}|p{40pt}|p{40pt}|p{50pt}|p{80pt}|}
\hline
\textbf{Pulse Period (s)}& 
\textbf{RA }& 
\textbf{Dec}& 
\textbf{Optical Counterpart}& 
\textbf{Name(s)} \\
\hline
0.087& 00 42 35 & -73 40 30& & AX J0043-737  \\
\hline
0.716 & 01 17 05.1& -73 26 35& B1Iab& SMC X-1  \\
\hline 
2.374 & 00 54 33 & -73 41 04& B1.5Ve& SMC X-2  \\
\hline
2.8& 00 59 12.9& -71 38 50& B0III-Ve& RX J0059.2-7138  \\
\hline
3.34 & 01 05 06& -72 11 08& -& AX J0105-722  \par ( RX J0105.3-7210?) \\
\hline
4.782& 00 53 07 & -72 17 24& B0V-B1Ve& XTE J0052-723  \\
\hline
7.77 & \multicolumn{2}{|p{96pt}|}{(AO7 Position A)} & -& XTE 7.77 \\
\hline
8.88 & 00 54 28.7& -72 45 37& B1e& 2E 0050.1-7247  \\
\hline
9.13 & 00 49 23 & -73 12 38& -& AX J0049-732  \\
\hline
15.3 & 00 52 15.5&-73 19 14& Be& RX J0052.1-7319  \\
\hline
16.6 & 00 51 51.2& -73 10 32& -& XTE \\
\hline
22.07 & 01 17 40 & -73 30 48& B1-2 III-Ve& RX J0117.6-7330  \\
\hline
31.0 & 01 11 14.5& -73 16 50& Be& XTE J0111.2-7317  \\
\hline
46.4& \multicolumn{2}{|p{96pt}|}{(Positions 4 {\&} 5)} & -& XTE 46.4 \\
\hline
46.6 & 00 53 56 & -72 26 54& Be& XTE  \par 1WGA J0053.8-7226 \\
\hline
51& \multicolumn{2}{|p{96pt}|}{(Positions 4 {\&} 5) } & -& XTE 51 \\
\hline
58.97 & 00 55 00 & -72 25 38& Be& XTE J0055-724, \par 1SAX J0054.9-7226,  \par 2E 0053.2-7242 \\
\hline
74.7 & 00 49 00 & -72 51 40& Be& XTE nameless,  \par AX J0049-729, \par RX J0049.1-7250 \\
\hline
82.4& 13.23& -72.23& -& XTE 82.4 \\
\hline
89& \multicolumn{2}{|p{96pt}|}{(AO7 Position A)} & -& XTE 89 \\
\hline
91.1 & 00 51 11 & -72 14 01& Be& XTE, AX J0051-722 \\
\hline
95.2 & \multicolumn{2}{|p{96pt}|}{AO4/5 POS1} & -& XTE SMC95  \\
\hline
101.4 & 00 57 27 & -73 25 31& -& AX J0057.4-7325  \\
\hline
169.3 & 00 54.6& -72 04& Be& XTE J0054-720  \par AX J0052.9-7157  \par RXJ0052.9-7158 \\
\hline
172.4 & 00 51 38 & -73 11 01& Be& AX J0051.6-7311 \\
\hline
280.4 & 00 57 58 & -72 02 56& & AX J0058-720  \\
\hline
323 & 00 50 47 & -73 15 51& Be& ASCA/ROSAT  \par RX J0050.8-7316 \\
\hline
348/343& 01 03 15 & -72 09 01& Be& AX J0103-722  \par 1SAX J0103.2-7209  \par 2E 0101.5-7225 \\
\hline
455+-2& 01 01 21 & -72 11 18& Be& RX J0101.3-7211 \\
\hline
755.5 & 00 49 33 & -73 23 23& & RX J0049.7-7323  \par AX J0049.4-7323 \\
\hline
\end{tabular}
\medskip

Pulsars in this table are ordered by pulse period. 
Spectral type of the optical counterpart is given if known. Where the 
source position is not accurately known, we give the monitoring positions 
(see Table ~\protect\ref{tab:positions}) in which the pulsar was detected.

\normalsize
\end{table*}

%% file: tab7.tex
\begin{table*}[htbp]
\caption{PCA collimator response $R$ for all SMC pulsars with well known positions.} 
\label{tab:responses}
\tiny
\begin{tabular}
{|p{20pt}|p{20pt}|p{20pt}|p{20pt}|p{20pt}|p{20pt}|p{20pt}|p{20pt}|p{20pt}|p{20pt}|p{20pt}|}
\hline
Pulsar& 
RA& 
Dec& 
Pos 1& 
Pos 2& 
Pos 3& 
Pos 4& 
Pos 5& 
Pos 1a& 
Pos 1b& 
Pos 1c \\
\hline
0.087s& 
10.65& 
-73.68& 
 & 
 & 
 & 
0.29& 
0.22& 
 & 
 & 
  \\
\hline
0.716s& 
19.25& 
-73.44& 
 & 
 & 
0.63& 
 & 
 & 
 & 
 & 
  \\
\hline
2.374s& 
13.64& 
-73.68& 
 & 
 & 
 & 
0.50& 
0.33& 
 & 
 & 
  \\
\hline
2.8s& 
14.80& 
-71.65& 
0.10& 
0.36& 
 & 
 & 
 & 
 & 
0.14& 
0.14 \\
\hline
3.34s& 
16.28& 
-72.19& 
0.11& 
0.91& 
 & 
 & 
 & 
 & 
 & 
0.18 \\
\hline
4.78s& 
13.08& 
-72.33& 
0.84& 
 & 
 & 
 & 
0.21& 
0.90& 
0.86& 
0.77 \\
\hline
8.88s& 
13.62& 
-72.76& 
0.68& 
 & 
 & 
0.42& 
0.53& 
0.63& 
0.41& 
0.68 \\
\hline
9.13s& 
12.35& 
-73.21& 
0.16& 
 & 
 & 
0.89& 
0.88& 
0.19& 
 & 
0.13 \\
\hline
15.3s& 
13.06& 
-73.32& 
0.12& 
 & 
 & 
0.88& 
0.73& 
0.11& 
 & 
0.10 \\
\hline
16.66s& 
12.96& 
-73.18& 
0.25& 
 & 
 & 
0.88& 
0.85& 
0.25& 
 & 
0.23 \\
\hline
22.07s& 
19.42& 
-73.51& 
 & 
 & 
0.54& 
 & 
 & 
 & 
 & 
  \\
\hline
31.0s& 
17.81& 
-73.28& 
 & 
 & 
0.67& 
 & 
 & 
 & 
 & 
  \\
\hline
46.6s& 
13.48& 
-72.45& 
1.00& 
 & 
 & 
0.15& 
0.29& 
0.86& 
0.69& 
0.93 \\
\hline
58.97s& 
13.75& 
-72.43& 
0.91& 
0.17& 
 & 
0.10& 
0.23& 
0.78& 
0.64& 
0.98 \\
\hline
74.7s& 
12.25& 
-72.86& 
0.45& 
 & 
 & 
0.57& 
0.75& 
0.51& 
0.35& 
0.39 \\
\hline
82.4s& 
13.23& 
-72.55& 
0.87& 
 & 
 & 
0.26& 
0.41& 
0.87& 
0.65& 
0.82 \\
\hline
91.1s& 
12.80& 
-72.23& 
0.71& 
 & 
 & 
 & 
0.13& 
0.79& 
0.99& 
0.65 \\
\hline
101.4s& 
14.36& 
-73.43& 
 & 
 & 
 & 
0.49& 
0.37& 
 & 
 & 
  \\
\hline
169s& 
13.65& 
-72.07& 
0.62& 
0.20& 
 & 
 & 
 & 
0.59& 
0.68& 
0.62 \\
\hline
172.4s& 
12.91& 
-73.18& 
0.24& 
 & 
 & 
0.89& 
0.86& 
0.24& 
 & 
0.22 \\
\hline
280.4s& 
14.49& 
-72.15& 
0.57& 
0.46& 
 & 
 & 
 & 
0.48& 
0.47& 
0.63 \\
\hline
323s& 
12.70& 
-73.26& 
0.15& 
 & 
 & 
1.00& 
0.83& 
0.16& 
 & 
0.13 \\
\hline
343s& 
15.81& 
-72.15& 
0.23& 
0.86& 
 & 
 & 
 & 
0.11& 
 & 
0.30 \\
\hline
455s& 
15.34& 
-72.19& 
0.38& 
0.71& 
 & 
 & 
 & 
0.26& 
0.21& 
0.45 \\
\hline
755s& 
12.39& 
-73.39& 
 & 
 & 
 & 
0.85& 
0.71& 
 & 
 & 
  \\
\hline
\end{tabular}
\medskip

The $RA$ \& $Dec$ (J2000) for each pointing position are given in Table~\protect\ref{tab:positions}. 
Blank entries indicate that the source was not in the field of view. These values 
were used in the reduction of pulsar monitoring data.

\normalsize

\end{table*}